\documentclass[prodmode,acmtompecs]{acmsmall} 

\usepackage[ruled]{algorithm2e}
\usepackage{algorithmic}
\renewcommand{\algorithmcfname}{ALGORITHM}
\usepackage{multirow}
\SetAlFnt{\small}
\SetAlCapFnt{\small}
\SetAlCapNameFnt{\small}
\SetAlCapHSkip{0pt}
\IncMargin{-\parindent}

\acmVolume{V}
\acmNumber{N}
\acmArticle{A}
\acmYear{YYYY}

\begin{document}

\title{Effective Handling of Urgent Jobs - Speed Up Scheduling for Computing Applications }
\author{YASH GUPTA and KAMALAKAR KARLAPALEM
\affil{\\International Institute of Information Technology, Hyderabad}}

\begin{abstract}
A queue is required when a service provider is not able to handle jobs arriving over the time. In a highly
flexible and dynamic environment, some jobs might demand for faster execution at run-time especially when the resources are limited 
and the jobs are competing for acquiring resources. A user might demand for speed up (reduced wait time) for some of the jobs
present in the queue at run time. In such cases, it is required to accelerate (directly sending the job to the server)
urgent jobs (requesting for speed up) ahead of other jobs present in the queue for an earlier completion of urgent jobs. Under the assumption
of no additional resources, such acceleration of jobs would result in slowing down of other jobs present in the queue. In this paper,
we formulate the problem of Speed Up Scheduling without acquiring any additional resources for the scheduling of on-line speed up requests posed by
a user at run-time and present algorithms for the same.\\
\hspace*{5 mm} We apply the idea of Speed Up Scheduling to two different domains
- Web Scheduling and CPU Scheduling. We demonstrate our results with a simulation based model using trace driven workload and 
synthetic datasets to show the usefulness of Speed Up scheduling. Speed Up provides a new way of addressing urgent jobs, provides a different evaluation 
criteria for comparing scheduling algorithms and has practical applications.
\end{abstract}



\acmformat{Yash Gupta and Kamalakar Karlapalem, 2013. Effective Handling of Urgent Jobs - Speed Up Scheduling
for Computing Applications.}

\begin{bottomstuff}
Author's addresses: International Institute of Information Technology, Hyderabad 500032 India.\\
Author's email addresses: yash.guptaug08@students.iiit.ac.in (Yash Gupta) and kamal@iiit.ac.in (Kamalakar Karlapalem).
\end{bottomstuff}

\maketitle

\section{Introduction}
One does not like to wait. In most of the cases, the number of service providers are
less compared to the number of jobs requiring service. In such cases, a queue is required
because the service provider is not able to handle jobs as soon as they arrive. There are
many practical scenarios where queuing cannot be avoided. For example, accumulation
of web requests in a queue arriving at web server, a queue of packets at routers, a queue of
processes waiting for CPU resources etc. In such situations, the choice of scheduling
policy determines mean wait time, mean queue length and other performance measures.\\
\hspace*{5 mm}In real life, job execution often requires human intervention in its execution. Further in a highly 
dynamic environment, a particular job may raise the requirement
for urgent attention, especially when the resources are limited. The user may demand for speed up 
(reduced wait time/faster execution) for some specific job present
in the queue at run time. In such a case, we need to accelerate/speed up such job 
(that requested for speed up based on user demand) ahead of other jobs present in the queue to the server for service. 
However, such acceleration would result in slowing down of other jobs which are present ahead of it in the queue since 
\textit{no additional resources are used}. Thus, the problem
that arises is how to schedule such speed up requests posed by the user so as to facilitate their earlier execution but at the same time ensuring less impact of slow down on the other jobs.\\
\hspace*{5 mm}In modern connected world, there is increasing prevalence of cloud computing which can dynamically allocate 
resources owing to an application need. A job seeking urgent attention can be served by dynamically allocating more resources to it.
However, resources are not infinite. Even in such scenarios, there could be high possibility of tremendous resource contention. It is not 
always feasible to provide additional resources to an urgent job either due to unavailability or high cost of resources. In
such cases, speed up quickly process all the urgent jobs without acquiring any additional resources.\\
\hspace*{5 mm}Consider an example of CPU Scheduling of different processes. The processes are usually 
categorized into different priority classes. All the processes of same priority class are usually scheduled using first come first serve or Round Robin strategy. 
In such a case, it is possible for a user to request speed up for a specific user process. However, the 
requested speed up is not urgent enough to move the process to higher priority class but the process 
needs to be accelerated with respect to those having same priority. Under the assumption of limited resources, 
some other processes within same priority class have to be delayed in order for one user to achieve requested speed up. \\
\hspace*{5 mm}The idea behind speed up is to achieve faster execution of jobs (requested by the user at run-time) without acquiring additional resources 
but with an overhead of some other jobs getting delayed. Yet we want to facilitate earlier execution of urgent jobs 
(whose delay might have severe consequences) without unnecessarily slowing down the other jobs. It is important to note that
it might not be possible to achieve speed up for all the jobs that requested for it as well as it might be the case that
while speeding up some jobs, even some of the jobs that requested for speed up were slowed down \cite{kafeza}.
\subsection{Background and Related Work} 
\hspace*{5 mm}Regarding scheduling in real time systems there is lot of work done for priority, 
job deadlines, data deadlines, precedence constraints etc. but in every case once the ordering is assigned and a job is
scheduled it cannot be altered. In our speed up problem execution needs to be adjusted at run time in order to 
respond to on-line user requests for speed up. The job priority is not determined according to a predefined scheduling policy assignment criterion 
like real time scheduling policies, but according to the user on-line requests for speed up posed at run-time.\\
\hspace*{5 mm}Various scheduling techniques have
already been proposed such as First Come First Serve, Shortest Job First, Priority Scheduling, EDF
(Earliest Deadline First), Multilevel Queue scheduling, Round Robin etc. These scheduling policies perform effectively in 
improving the performance of the system in the absence of on-line speed up requests, but when such requests occur we need more
fine grained scheduling policies which can speed up all the activities that requested for it at run time. We can compare and
contrast different scheduling algorithms based on speed up/slow down characteristics. The results and analysis for such 
comparisons are presented and explained in later sections of this paper. When a specific job completes before 
its expected execution time with respect to a specific scheduler, then we refer this as speeding up of job. Shortest Job
First speeds up the job with shorter duration time, Priority scheduling speeds up the job with highest priority both with
respect to FCFS scheduler. However these scheduling policies are unable to handle on-line speed up requests
where execution needs to be adjusted at run time to provide remedies for the delayed and urgent jobs. Further existing scheduling policies speed up specific kind of jobs which do not take into consideration the slow down for other
jobs. In our problem of speed up, we might have to speed up even the delayed non-urgent jobs under certain circumstances.\\
\hspace*{5 mm}Earliest Deadline First (EDF) scheduling policy focuses on speeding up of jobs with earliest deadlines so that they do not
miss their deadlines. EDF scheduling problem seems similar to speed up problem in a sense that
it can be executed on-line, the priority depends on the jobs that arrive (specific deadlines) and the priority assignment
is done at run-time but there is a difference between preserving deadlines and problem of speed up. \textit{The notion
of deadline is absolute whereas concept of speed up is relative}. The deadlines of arriving jobs are absolute and independent of the state of the queue at their arrival whereas the concept of speed up is defined with respect to
expected execution time which is influenced by the queue state. In EDF, the 
deadlines are met by speeding up of urgent jobs but in our case, there need not be any deadline to achieve speed up. EDF tries to
minimize the number of jobs that miss their deadlines, whereas Speed Up focus is to
accelerate/speed up as many jobs as possible which requested for it at run-time. Further, EDF scheduling
does not care about the jobs which do not have specific deadlines (non urgent), but in our case we 
cannot penalize more than necessary to the rest of the executing jobs. In speed up problem, it is more beneficial to 
achieve the exact requested speed up for urgent job rather than achieving relatively high amount of speed up (than requested)
thereby causing higher amount of slow down to the rest of the executing jobs.\\
\hspace*{5 mm} Speed Up provides a new way of addressing urgent jobs, provides a different evaluation 
criteria for comparing scheduling algorithms and has practical applications. There has been little work on Speed Up 
scheduling except \cite{wecwis}, \cite{kafeza2000speeding}, \cite{kafeza2001speeding} and \cite{kafeza}.
The authors modeled speed up problem for speeding up of E-commerce activities \cite{wecwis} and improving the response time of 
business processes \cite{kafeza}. The authors presented three explicit speed up algorithms (MPF, MPF-SD and MinPF) 
based on location table model where they select a specific job based on its queue position and achieve speed up by swapping of jobs in the queue.
MPF (Maximum Position First) algorithm tries to push forward all the jobs requesting for speed up at the head of the queue and decides their
order by trying to swap small positions (near to the head) with the large ones (end of the queue). MPF-SD (MPF-Smaller Duration) 
algorithm is similar to MPF algorithm but it preserves one property that ``\textit{
no job that is requesting for speed up is delayed}''. The urgent job (requesting for speed up) is pushed ahead in the queue
if and only if its duration is smaller than the one with which it swaps. MinPF (Minimum Position First) algorithm is compliment of MPF algorithm.
Instead of swapping jobs with maximum distance, it swaps jobs with minimum distance. All the three algorithms
are purely positional and computationally expensive with an overhead of maintaining a location table. Further, authors in \cite{kafeza}
and \cite{wecwis} used only \textit{achieved ratio} metric (percentage of jobs that achieved their requested speed up)
for evaluating the effectiveness of their speed up algorithms. Their work did not consider the impact of slow down
on the remaining jobs as a result of speed up which is crucial. Any scheduling algorithm speeding up some urgent jobs requiring
urgent attention but causing arbitrarily high slow down to the rest of the jobs would not be practical and fair. Therefore along with 
achieved ratio, we also consider the impact of slow down to the 
non urgent jobs as a consequence of speed up and the overall mean wait time. In this paper, we remodel the speed up problem aiming at speeding up the jobs which requested for it 
 but at the same time providing less scope of slow down for the rest of the executing jobs while keeping the mean wait time
 reasonable. We provide implicit techniques to address speed up problem where the notion of acceleration is incorporated in the priority function, thus
leading to computationally efficient solution.\\
\hspace*{5 mm}The rest of the paper is structured as follows. The Speed Up scheduling problem
is formulated in Section 2 and Speed Up algorithms are presented in Section 3. Section 4 
presents the experimental analysis of our proposed algorithms. Section 5 and 6
respectively describes about the applications of Speed Up in web scheduling and CPU
scheduling. Section 7 concludes our work along with possible future work.

\section{Speed Up Problem}
\hspace*{5 mm}Throughout this work, a single server queuing model is used for addressing the problem of speed up. Even in the presence
of multiple queues, we need to support speed up within every single queue associated with a specific server. 
Load balancing (movement of jobs across different
queues to equally distribute workload on multiple servers) is a separate problem which may or may not help in achieving speed up (when the queues are equally loaded).
Supporting speed up through load balancers is beyond the scope of this work.\\
\hspace*{5 mm}Consider jobs arriving at a system with respect to time. The system comprises of
a single server which can serve at most one job at any point of time and a queue
where the jobs accumulate and wait for their service. \textit{The jobs are assumed to be non pre-emptable in this 
work}. The job once finished leaves the system. The state of the
system is a snapshot of the queue at a particular instant of time.
\begin{definition}[System State]
State of the system at any point of time $t$ denoted by $S(t)$, is defined as the tuple $<j_{1},j_{2},j_{3},.....,j_{l}>$
where $l$ is the queue length at time $t$ and $j_{i}$ is the job at $ith$ position waiting in the queue for service,
such that $1 \le i \le l$ and each job $j_{i}$ having arrival time strictly less than $t$. The job getting its service
at $t$ is not included in $S(t)$.
\end{definition}
\hspace*{5 mm}The concept of speed up by itself is relative \cite{kafeza}. Therefore, we define an
expected execution time for each job $j$ and then define speed up with respect to it.
The expected execution time for job $j$ (expected total time spent in the system from its arrival till it is completely
finished) with respect to scheduling policy $P$ is defined as the summation of the processing time of all the jobs that would be executed before $j$ depending upon
$P$ plus the processing time of job $j$. Note that the processing time of job 
is the service time of job also referred to as duration of job. If $P$ scheduling policy is assumed to be FCFS, then expected execution time for job $j$ would be defined as summation of processing time of all the jobs
present in the queue which arrived before $j$ plus the processing time of $j$. \textit{Note that, the definition of expected execution time would change if some other
scheduling technique is assumed}. Throughout this work, we use the term expected execution time with respect to FCFS scheduler.
\begin{definition}[Expected Execution Time]
\label{expecteddef}
The expected execution time for job $j$ with respect to FCFS scheduler arriving at time $t$, denoted by 
$T_{exp}(j)$ is defined as \begin{equation}T_{exp}(j) = d(j) + \sum\limits_{\forall k \in S(t)} d(k) + d_{rem}(j_{s}),\end{equation} 
where $d(k)$ denotes the processing time/duration for job $k$ and $d_{rem}(j_{s})$ denotes the remaining processing time of job $j_{s}$ 
being served at the server. If there is no job at server being served at $t$, then value of $d_{rem}(j_{s})$ is zero.
\end{definition}
\begin{definition}[Actual Execution Time]
The actual execution time for job $j$, denoted by $T_{actual}(j)$ is the actual total time spent 
 by the job in the system from its arrival time till it leaves the system (completely finishes).
 \begin{equation}
  T_{actual}(j)= T_{finish}(j) - T_{arrival}(j)
 \end{equation}
 where $T_{finish}(j)$ and $T_{arrival}(j)$ respectively denotes the finish time (time at which job $j$ completes its
 execution) and arrival time for job $j$.
\end{definition}
\hspace*{5 mm}Note that the actual execution time for job can be computed only after the job finishes.
\begin{definition}
A job $j$ is said to achieve speed up iff $T_{exp}(j) - T_{actual}(j) > 0$ and is said to
be slowed down iff $T_{exp}(j)-T_{actual}(j) < 0$.
\end{definition}
\hspace*{5 mm}Note that the jobs for which their expected execution time is equal to their actual execution time, are neither speeded
up nor slowed down. If first come first serve scheduling would have been used, then actual execution time will be equal to expected execution
time for all of the jobs and thus, all the jobs will neither be speeded up nor slowed down. Depending on the values of expected and actual execution time
we can say whether a specific job is speeded up or slowed down. \\
\hspace*{5 mm}We dynamically label some of the arriving jobs as urgent (\textit{i.e.} they are requesting 
speed up). The objective of Speed Up problem is to speed up all the jobs that are requesting for speed up
without penalizing more than necessary to the rest of the jobs present in the queue, while keeping the mean wait time low
for all the jobs. It might not be possible to achieve speed up for all jobs which requested for it. For example, consider a scenario where all 
the jobs present in the queue are requesting for speed up, then it is impossible to achieve speed up for all. Also
it might be the case that in the process of speeding up of some of the urgent jobs, even some of the jobs that requested for speed up
were slowed down. This case may arise when an urgent job having high processing time is accelerated ahead of another urgent job
in the queue.\\
\hspace*{5 mm}Consider a queue of jobs as $<j_{1}, j_{2}, j_{3}, ....., j_{l} >$ present in the system at some point of time $t$ 
where $l$ is the queue length at $t$. Suppose job $j_{i}$ for some $1 < i \le l$ has requested for speed up. In order to achieve speed up for job $j_{i}$, it needs to be
accelerated to the server. This acceleration will have no impact on all the jobs $j_{k}$ where $i+1 \le k \le l$. But all jobs 
$j_{p}$ such that $1 \le p \le i-1$ ahead of $j_{i}$ will suffer a delay equal to $d(j_{i})$ (duration time of job $j_{i}$). 
 Let there be an urgent job
$j_{k}$ for some $k \in [1,i-1]$ which is temporarily delayed because of the acceleration of $j_{i}$, but it is possible that the job $j_{k}$ later got accelerated so that eventually it achieved speed up despite of facing an initial delay. 
\textit{Therefore it is not always the case, that a delayed job will be slowed down by the time it completes its execution}. Thus, 
the priority/importance of any job at any point of time $t$ dynamically changes depending on how much it has been delayed till $t$ 
because of acceleration of other jobs. The scheduling decisions need to be made according to the on-line speed up requests posed by the user at run-time.
\begin{definition}
A job $j$ is said to achieve speed up of $x$ time units iff $T_{exp}(j)-T_{actual}(j)
= x$ and $x>0$ and is said to be slowed down by $z$ time units iff $T_{actual}(j) - T_{exp}(j) = z$ and $z>0$. 
\end{definition}
\begin{table}
\centering
\tbl{Information of jobs arriving to the system.\label{info}}{
\setlength{\tabcolsep}{10pt}
\begin{tabular}{|l|l|l|} \hline 
Job Id&Arrival Time& Job Duration\\ \hline
$j_{0}$&0&5\\ \hline
$j_{1}$&1&5\\ \hline
$j\prime_{2}$&2&3\\ \hline
$j\prime_{3}$&3&4\\ \hline
$j_{4}$&4&3\\ \hline
$j\prime_{5}$&5&5\\ \hline
\end{tabular}}
\end{table}
\begin{example}
Consider the jobs arriving as per Table \ref{info}. Let jobs $j\prime_{2}$, $j\prime_{3}$ and $j\prime_{5}$ are the jobs
which are requesting for speed up. Suppose the order of schedule is {$j_{0}, j\prime_{5}, j\prime_{3}, j\prime_{2}, j_{4}, j_{1}$}.
First job $j_{0}$ is picked by the server for service at time $t = 0$. At $t=0$ since there is no job in the queue therefore, $T_{exp}(j_{0})$ is $5$
equal to its own duration time. Thus, job $j_{0}$ is neither speeded up nor slowed down. At $t=1$, job $j_{1}$ arrives and it finds
that there is no job in the queue waiting for its service and $j_{0}$ is being served by the server, therefore 
$T_{exp}(j_{1})$ is $5+4$ (remaining service time of $j_{0}$) that is 9 time units. At $t=2$, 
job $j\prime_{2}$ finds job $j_{1}$ waiting in the queue, and thus $T_{exp}(j\prime_{2})$ is $3$ (d($j\prime_{2}$)) $+$ $5$ (d($j_{1}$)) $+ $ $3$ (remaining service time of $j_{0}$) that is 11 time units. Similarly, expected execution time values 
for $j\prime_{3}$, $j_{4}$ and $j\prime_{5}$ are respectively $14$, $16$ and $20$ units. At $t=5$, when $j_{0}$ finishes, 
urgent job $j\prime_{5}$ is speeded up to the server for service. Then all the jobs from $j_{1}$ to $j_{4}$ will be delayed by 5 time units including the urgent
jobs $j\prime_{2}$ and $j\prime_{3}$. The job $j\prime_{5}$ finishes at $t$ = $10$. The actual execution time for $j\prime_{5}$ is thus, 5 units. Since 
$T_{exp}(j\prime_{5}) > T_{actual}(j\prime_{5})$, $j\prime_{5}$ is said to achieve speed up of 20-5 = 15 units. Now the delayed job $j\prime_{3}$ is selected for service
after $j\prime_{5}$ at $t = 10$. Then $j\prime_{3}$ will finish at $t = 14$. The $T_{actual}(j\prime_{3})$ is thus $14-3 = 11$ time units. However,
the expected execution time for $j\prime_{3}$ was 14 units. Thus even after facing the initial delay, $j\prime_{3}$ is successful in 
achieving speed up of $3$ time units. Now job $j\prime_{2}$ will be scheduled next and will finish at $t=17$. The actual execution
time for $j\prime_{2}$ is thus 17-2 = 15 units. However, the expected execution time for $j\prime_{2}$ is 11 units. Thus, job
$j\prime_{2}$ is slowed down by 15-11 = 4 units despite of requesting for speed up. Similarly job $j_{4}$ followed by $j_{1}$ executes.
The information about expected execution time, actual execution time, speed up/slow down for this example as per the supposed schedule is shown in Table \ref{result}.\\
\end{example}
\begin{table}
\tbl{Speed Up/Slow down information.\label{result}}{
\setlength{\tabcolsep}{4pt}
\begin{tabular}{|l|l|l|l|l|l|} \hline 
Job Id&Arrival Time&Job Duration&$T_{exp}$&$T_{actual}$&Speed Up/Slow Down \\ \hline
$j_{0}$&0&5&5&5&no speed-up/slow-down\\ \hline
$j_{1}$&1&5&9&24&slow down by 15 units\\ \hline
$j\prime_{2}$&2&3&11&15&slow down by 4 units\\ \hline
$j\prime_{3}$&3&4&14&11&speed up by 3 units\\ \hline
$j_{4}$&4&3&16&16&no speed-up/slow-down\\ \hline
$j\prime_{5}$&5&5&20&5&speed up by 15 units\\ \hline
\end{tabular}}
\end{table}
\subsection{Problem Classification}
The problem of speed up can be classified into following two categories depending upon whether the urgent jobs are
requesting for an exact amount of speed up or not :
\begin{enumerate}
 \item \textit{Speed Up with no constraints:} In this problem, some of the jobs present in the queue are requesting speed
 up, however there is no constraint on the amount of speed up that they request. The purpose is to achieve speed up for
 as many urgent jobs as possible without unnecessarily slowing down the other jobs while keeping the mean wait time low
 for all the jobs.
 \item \textit{Fine Grained Selective speed up:} In this problem, the urgent jobs request speed up of some specific 
 time units. The maximum amount of speed up that a job $j$ can achieve is clearly equal to the summation of duration times 
 of all the jobs that are in front of it at the time of it's arrival. So each urgent job can request for some percentage of the maximum speed up
 that it can achieve at that time. Formally, amount of speed up requested by an urgent job $j$ denoted by $RSU(j)$ arriving at time $t$, 
 is given by
 \begin{equation}\label{RSUeq} RSU(j) = (\frac{p_{j}}{100}) \times \sum\limits_{\forall k \in S(t)} d(k)\end{equation} for some $p_{j} \in (0,100]$ where $RSU(j)$ stands 
 for \textit{Requested Speed Up} by job $j$, $S(t)$ represents the system state
 at time $t$ and $d(k)$ is the duration time of job $k$.\\
 \hspace*{5 mm}The objective of Fine Grained Selective Speed Up is to achieve (almost) exact 
 requested speed up for
 as many urgent jobs as possible without unnecessarily slowing down other jobs while keeping the mean wait time low for all the 
 jobs.
 \end{enumerate}
\hspace*{5 mm}To summarize, Speed Up problem aims at meeting the following objectives:
\begin{itemize}
 \item To maximize the number of urgent jobs which achieve speed up (exact requested speed up in case of fine grained selective speed up).
 \item Avoid unnecessary slow down for other jobs while accelerating urgent jobs.
 \item Reducing the overall mean wait time for all the jobs.
 \item Avoid starvation for any of the job (whether urgent or non-urgent).
\end{itemize}

\section{Speed Up Algorithms}
\hspace*{5 mm}We provide implicit techniques where the notion of acceleration is incorporated in the priority function. We present following three strategies for addressing speed up problem depending on the amount of slow down
that can be tolerated for the non-urgent jobs:
\begin{enumerate}
 \item \textit{Urgent Dominant Speed Up (UDSU):} The focus of this technique is to accelerate the urgent jobs as much as possible 
 without even bothering about the slow down caused to other non-urgent jobs while keeping the mean wait time low. This strategy 
 gives very high priority to the urgent jobs and might even starve non-urgent jobs in the presence of continuous arrival of urgent
 jobs.
 \item \textit{Non-Urgent Benevolent Speed Up (NUBSU):} This strategy gives high priority to the urgent jobs, but at the same time 
 it opportunistically tries to accelerate delayed non urgent jobs as long as such acceleration is not a hurdle 
 in achieving speed up for rest of the urgent jobs present in the queue at that time. 
 \item \textit{Fair Speed Up (FSU):} This strategy is fair to all the jobs and does not lead to high slow down for
 any of the job. FSU strategy is neither biased towards urgent jobs nor does create starvation for any job while 
 at the same time aiming at reducing the overall mean wait time. With this strategy we can still implicitly achieve speed up
 for some jobs.
\end{enumerate}
\hspace*{5 mm}The priority function for job $j$, denoted by $P(j)$ is defined as follows:
\begin{equation}
\label{Pfunction}
 P(j) = T_{arrival}(j) * d(j)
\end{equation} where $T_{arrival}(j)$ denotes the arrival time of job $j$ and $d(j)$ denotes it's duration time.\\
\hspace*{5 mm}The job with lower value of priority function is preferred. The idea is to give preference to small size jobs 
(to improve mean wait time) but this preference
takes into consideration the arrival time of job to avoid starvation. The priority function ensures no starvation for jobs 
while still preferring shorter jobs (to improve mean wait time) 
because for each job $j$, since job sizes are finite,
there exists certain arrival time (for some other job $j\prime$) $T_{arrival}(j\prime)>T_{arrival}(j)$
such that all the jobs $j\prime$ after $j$ (irrespective of their sizes)
will have higher priority value than $j$. Thus, each job $j$ would
eventually be served.\\
\hspace*{5 mm}\textit{Though the priority function has some commonality with existing scheduling techniques, but we study it in context of
speed up and slow down. Further, we extend our priority function for different applications (refer Section 5 and 6).}
\begin{definition}[Candidate Job]
Let $J(t)$ be any set comprising of jobs present in the system at time $t$ including
the jobs that arrive at $t$. A job $j$ is 
said to be \textit{Candidate} with respect to set $J(t)$ for some function $p(j)$ denoted by $C_{J,p}(t)$, iff $j \in J(t)$ and 
$p(j)$ is minimum 
among all the jobs present in $J(t)$. If there are
multiple such jobs with minimum function value, then $C_{J,p}(t)$ is the one with shortest duration time among them.
\end{definition}
\hspace*{5 mm}At any point of time $t$, let $J_{urgent}(t)$ and $J_{nonurgent}(t)$ respectively denote the set of all the urgent
(jobs requesting for speed up) and
non-urgent jobs waiting in the queue for service.
\subsection{Speed Up with no constraints}
\hspace*{5 mm}In this problem, some of the jobs present in the queue are requesting for speed up. However, there is no constraint on the
amount of speed up requested. The objective is to achieve speed up for urgent jobs (actual execution time should be less than
expected execution time), without penalizing more than necessary to the rest of the jobs.
\subsubsection{Urgent Dominant Speed Up Algorithm (UDSU)}
\begin{algorithm}
\caption{UDSU algorithm.}
\textbf{Input:} Input queue of jobs (queue size $\ge 1$) at time $t$.\\*
\textbf{Output:} Next job to be executed.\\
\begin{algorithmic}[1] \label{algo:blind-1}
\STATE $j \gets $ \textit{null}.
\IF{ $J_{urgent}(t) = \phi$ }
\STATE $j \gets C_{J_{nonurgent},P}(t) $ \\
// $T_{arrival}(j)*d(j)$ is minimum and $j \in J_{nonurgent}(t)$
\ELSE
\STATE $j \gets C_{J_{urgent},P}(t)$ \\
// $T_{arrival}(j)*d(j)$ is minimum and $j \in J_{urgent}(t)$
\ENDIF
\RETURN $j$
\end{algorithmic}
\end{algorithm}
UDSU Algorithm accelerates the urgent jobs to the server based on the priority value of it.
The urgent job with least $P$ (as defined in \ref{Pfunction}) function value is chosen (in case of ties the one with shorter duration time). If there is no urgent 
job present in the queue, then the non-urgent job with least $P$ priority function value is chosen. The notion of acceleration is incorporated
in the priority function. This algorithm completely ignores the non-urgent jobs in the presence of 
urgent jobs and focuses on accelerating as many urgent jobs as possible, thus referred to as \textit{Urgent Dominant Speed
Up} algorithm. Note that this algorithm will cause the other non-urgent jobs to have arbitrarily high slow down or
can even create starvation for them in the presence of continuous arrival of urgent jobs.\\
\hspace*{5 mm} To implement this algorithm, we need to maintain two separate \textit{priority queues} one for the urgent and the other 
one for non-urgent jobs, based on their priority value. The time complexity of UDSU algorithm to pick up 
the next job at time $t$ is  $O(log($ $\left|{J_{urgent}}(t)\right|)$ $+$ $log($ $\left|{J_{nonurgent}}(t)\right|)$ $)$ and is quite efficient
as compared to all the three existing algorithms which have time complexity $O(\left|J_{urgent}(t)\right| + \left|J_{nonurgent}(t)\right|)$.
\subsubsection{Non-Urgent Benevolent Speed Up Algorithm (NUBSU)}
This algorithm gives high priority to the jobs which are requesting speed up, but is not unfair towards other jobs. It
optimistically tries to accelerate other jobs as long as such acceleration is not posing any hurdle in achieving speed up
for existing urgent jobs.
\begin{definition}[Opportunistically forwardable job]
A job $j$ is said to be opportunistically forwardable at time $t$
 if $j \in J_{nonurgent}(t)$ and $\forall k \in J_{urgent}(t)$,
\begin{equation}[T_{wait}(k,t)+d(j)+ \newline
\sum\limits_{P(k\prime) \le P(k)} d(k\prime) + d_{rem}(j_{s})] < T_{exp}(k)\end{equation} where $k\prime \in J_{urgent(t)}$ such that $P(k\prime) \le P(k)$
 , $T_{wait}(k,t)$ denotes the waiting time of job $k$ till time $t$ and $d_{rem}(j_s)$ is the remaining 
 duration time of job $j_s$ getting serviced at server at $t$ if any else $zero$.
\end{definition}
\begin{algorithm}
\caption{NUBSU algorithm}
\textbf{Input:} Input queue of jobs (queue size $\ge 1$) at time $t$.\\*
\textbf{Output:} Next job to be executed.\\
\begin{algorithmic}[1]
\label{algo:opp-1}
\STATE $j \gets $ \textit{null}.
\IF{ $J_{nonurgent}(t) = \phi$ \textbf{or} $C_{J_{nonurgent},P}(t)$ is \textbf{not} \textit{opportunistically forwardable} at $t$}
\STATE $j \gets C_{J_{urgent},P}(t)$ \\
// $T_{arrival}(j)*d(j)$ is minimum and $j \in J_{urgent}(t)$
\ELSE
\STATE $j \gets C_{J_{nonurgent},P}(t)$ \\
// $T_{arrival}(j)*d(j)$ is minimum and $j \in J_{nonurgent}(t)$
\ENDIF
\RETURN $j$
\end{algorithmic}
\end{algorithm}                                                         
\hspace*{5 mm}A non urgent job $j_{nonurgent}$ is opportunistically forwardable at time $t$, if it's acceleration 
to the server for service will not affect the existing urgent jobs (present in the queue) in achieving their speed up at $t$. That is, existing urgent jobs can achieve 
speed up even after the acceleration of non-urgent job $j_{nonurgent}$ to the server for service. In such case, each 
urgent job $j_{urgent}$ has to further wait for all the other urgent jobs which have lower $P(j)$ 
function value (higher preference) 
than $j_{urgent}$ plus the duration time of $j_{nonurgent}$ since $j_{nonurgent}$ is opportunistically forwardable and thus, will be 
accelerated for service before $j_{urgent}$. Thus, for $j_{urgent}$ to achieve speed up, the time that it has to further
wait (including it's own duration time) plus the time that it has already spent waiting in queue till now should be less than it's expected execution time.\\
\hspace*{5 mm} NUBSU Algorithm gives high priority to the urgent jobs, but is still optimistic 
towards non urgent jobs in a sense that it accelerates them whenever such acceleration does not pose any hurdle for urgent jobs in 
achieving their speed up. The important question is does acceleration of opportunistically forwardable job always ensure that the other urgent jobs 
will achieve speed up? The answer is no. Our algorithms are on-line and we do not have any information about the jobs arriving in
future. So it may happen that at time $t$, a non-urgent job $j_{nonurgent}$ is opportunistically forwardable and thus is 
forwarded to the server for service, but during
the execution of $j_{nonurgent}$, a bunch of urgent jobs arrives in the queue say at time $t+\delta t$ and some of which are having their $Priority$ function value
lower than some existing urgent job $j_{urgent}$ (might be because of very low duration time). So $j_{urgent}$ for which summation of duration times of all the urgent jobs having
lower $Priority$ function value plus duration of $j_{nonurgent}$ was less than $T_{exp}(j_{urgent})$ at time $t$, can now become greater than
$T_{exp}(j_{urgent})$ at time $t+\delta t$ (because of arrival of some urgent jobs which would have been executed before $j_{urgent}$ because of their
low $P$ function value). Thus, $j_{urgent}$ cannot achieve speed up. This situation is analogous to non preemptive SJF scheduling
where recently arrived short job has to wait for currently executing long job to finish (since the algorithm is on-line and 
non preemptive). To measure the probability of such events, we define the notion
of \textit{successful/unsuccessful} opportunistically forwarded non urgent jobs. A \textit{successful opportunistically forwarded non-urgent job} 
$j_{nonurgent}$ is the one whose acceleration to the server for service does not impact any other existing urgent job in 
achieving speed up by the time it complete its execution. Similarly, \textit{unsuccessful opportunistically forwarded job} $j_{nonurgent}$ is the one whose acceleration 
prevented some of the existing urgent jobs in achieving their speed up. 
\begin{definition}[Degree of Non Urgent Job Impact (DNUJI)]
Degree of Non Urgent Job Impact (DNUJI) for NUBSU algorithm is defined as the ratio of 
number of unsuccessful opportunistically forwarded jobs to the total number of opportunistically forwarded jobs.\\
\begin{equation}
 DNUJI = \frac{NumberOfUnsuccessfulOpportunisticallyForwardedJobs}{TotalNumberOfOpportunisticallyForwardedJobs}
\end{equation}
\end{definition}
\hspace*{5 mm}To implement NUBSU algorithm, we maintain a priority queue for non urgent jobs based on $P$ function value.
For urgent jobs, we maintain a AVL tree based on their $P$ function value. To check whether $C_{J_{nonurgent},P}(t)$
is {opportunistically forwardable}, we need to traverse the AVL tree in increasing order of $P$ function values by doing an
in-order traversal of tree and checking for each urgent job $k$, whether
\begin{equation}[T_{wait}(k,t)+d(C_{J_{nonurgent},P}(t))+ \newline
\sum d(k\prime) + d_{rem}(j_s)] < T_{exp}(k)\end{equation} where $k\prime \in J_{urgent(t)}$ such that $P(k\prime) \le P(k)$
and $d_{rem}(j_s)$ denotes the remaining duration time of currently executing job $j_s$.
The time complexity for obtaining $C_{J_{urgent},P}(t)$ and $C_{J_{nonurgent},P}(t)$ is 
$O(log($ $\left|{J_{urgent}}(t)\right|)$ $+$ $log($ $\left|{J_{nonurgent}}(t)\right|)$ $)$. However, the time complexity of determining whether $C_{J_{nonurgent},P}(t)$ is opportunistically forwardable 
is $O($ $\left|{J_{urgent}}(t)\right|)$ since we need to traverse the whole AVL tree containing urgent jobs at time $t$. Thus, the 
overall time complexity of NUBSU Algorithm for picking up the next job at time $t$ is 
$O(log($ $\left|{J_{urgent}}(t)\right|)$ $+$ $log($ $\left|{J_{nonurgent}}(t)\right|)$ $+$ $\left|{J_{urgent}(t)}\right|$  $)$. 
This algorithm is computationally more expensive than UDSU algorithm.
\subsubsection{Fair Speed Up Algorithm (FSU)}
This algorithm is not biased to any specific kind of job. It chooses the job which is
having least $P$ function value among all the jobs present in the queue (including both urgent and non-urgent jobs) to 
schedule next. If there are multiple such jobs, then the job with smaller duration time is selected. Note that, both UDSU
as well as NUBSU algorithms can create starvation for non-urgent jobs
in the presence of continuous arrival of urgent jobs. However, FSU provides less scope of starvation for any of the job. Each
job would eventually be served because the next job will be the one having least $P$ function value among all the jobs present
in the queue. Therefore, a job waiting for too long will have low value of arrival time as compared to other jobs which arrived
after it, and since job sizes are finite, thus a time $t$ will always exist such that its $P$ function value is minimum among all the jobs in the system. 
The major drawback of this algorithm is that it does not
serve the purpose of achieving speed up for as many urgent jobs as possible, but we can still
implicitly achieve speed up for some of the jobs. The benefit of this algorithm is that it does not starve any of the job
along with improving the overall mean wait time.\\
\hspace*{5 mm}To implement this algorithm, we need only one \textit{priority queue} containing all the jobs present in the
queue based on their $P$ priority value. Thus, the time complexity to pick up the next job for FSU algorithm is $O(log$ $l$ $)$ where $l$ is the length of the queue.
\subsection{Fine Grained Selective speed Up}
\hspace*{5 mm}In this speed up problem, each of the urgent job requests for a specific amount of speed up denoted by $RSU(j)$
which from eq. \ref{RSUeq} is defined as
\begin{equation} RSU(j) = \frac{p_{j}}{100} \times (\sum\limits_{\forall k \in S(T_{arrival}(j))} d(k))\end{equation} for some $p_{j} \in (0,100]$.\\
\hspace*{5 mm}By the definition \ref{expecteddef} of $T_{exp}(j)$, $RSU(j)$ can also be written as
\begin{equation}\label{newRSUeq} RSU(j) = \frac{p_{j}}{100} \times ( T_{exp}(j)-d(j)-d_{rem}(j_s) ) \end{equation}
\begin{definition}
An urgent job $j_{urgent}$ is said to be \textit{successfully speeded up} iff 
\begin{equation}
(T_{exp}(j_{urgent}) - T_{actual}(j_{urgent}) ) \ge RSU(j_{urgent})
\end{equation}
\end{definition}
\hspace*{5 mm}The objective of Fine Grained Selective Speed Up is to maximize the number of \textit{successfully speeded up} jobs without
unnecessarily slowing down other jobs while keeping the mean wait time low.
\begin{definition}
The current speed up for an urgent job $j_{urgent}$ at time $t$, denoted by
$CSU(j_{urgent},t)$ is defined as follows:
\begin{equation}\label{CSUeq} CSU(j_{urgent},t) = T_{exp}(j_{urgent}) - ( T_{wait}(j_{urgent},t) + d(j_{urgent}) + d_{rem}(j_s)) \end{equation}
where \begin{equation}\label{waiteq}
       T_{wait}(j_{urgent},t) = t - T_{arrival}(j_{urgent})
      \end{equation}
\end{definition}
\hspace*{5 mm}The current speed up at point of time $t$ for job $j$ represents the amount of speed up that it will achieve if it 
would have been accelerated next to the server for service just after the execution of job currently being served.\\
\hspace*{5 mm} The priority of an urgent job $j$ at time $t$ denoted by $P_{urgent}(j,t)$ is defined as:
\begin{equation} P_{urgent}(j,t) = CSU(j,t) - RSU(j)\end{equation}
Using eq. \ref{newRSUeq}, \ref{CSUeq} and \ref{waiteq}, we get
\begin{equation} P_{urgent}(j,t)=[T_{arrival}(j)+(1-p_{j}/100)*(T_{exp}(j)-d(j)-d_{rem}(j_s))] - t \end{equation} 
\hspace*{5 mm}However, the time $t$ at which we will calculate the priority value and $d_{rem}(j_s)$ remains the same for all urgent jobs, so value of $t$ 
and $d_{rem}(j_s) $ won't affect
 in comparing the priorities and thus, we can ignore it. Therefore,
\begin{equation}\label{Purgenteq} P_{urgent}(j) = T_{arrival}(j)+(1-p_{j}/100)*[T_{exp}(j)-d(j)] \end{equation}
\hspace*{5 mm}The function $P_{urgent}$ is used as priority function among urgent jobs (lower the value of function, higher the
importance/preference of it). It can be inferred from eq.\ref{Purgenteq} that a job with low arrival time requesting for
high percentage of speed up having low expected execution time is preferred. Note that, an urgent job $j$ will be successfully speeded up iff $CSU(j,t) \ge RSU(j)$ 
where $t$ is the time at which job $j$ was accelerated to the server for service. For non-urgent jobs, we use previous 
priority function $P$ defined in eq. \ref{Pfunction} for computing their priority.\\
\hspace*{5 mm}In order to present the three strategies for speed up for this case, we present a 
\textit{Generic Probabilistic Speed Up (GPSU)} 
algorithm which based on the value of probability $p$ can mimic UDSU, NUBSU and FSU algorithm. 
\subsubsection{Generic Probabilistic Speed Up Algorithm (GPSU)}
\begin{algorithm}
\caption{GPSU Algorithm.}
\textbf{Input:} Input queue of jobs (queue size $\ge 1$) at time $t$ and probability parameter $p$\\*
\textbf{Output:} Next job to be executed.\\
\begin{algorithmic}[1]
\label{algo:prob}
\STATE $j \gets $ \textit{null}.
\IF{ $J_{urgent}(t) = \phi$ }
\STATE $j \gets C_{J_{nonurgent},P}(t)$
\RETURN $j$
\ENDIF
\IF{ $J_{nonurgent}(t) = \phi$ }
\STATE $j \gets C_{J_{urgent},P_{urgent}}(t)$
\RETURN $j$
\ENDIF
\STATE Generate a random number $x$ between 0 to 1 inclusive.
\IF { $x \le p$ }
\STATE $j \gets C_{J_{urgent},P_{urgent}}(t)$ \\
// $P_{urgent}(j)$ is minimum and $j \in J_{urgent}(t)$
\ELSE
\STATE $j \gets C_{J_{nonurgent},P}(t)$ \\
// $T_{arrival}(j)*d(j)$ is minimum and $j \in J_{nonurgent}(t)$
\ENDIF
\RETURN $j$
\end{algorithmic}
\end{algorithm}
GPSU Algorithm takes a probability parameter $p$ which is a measure of bias between urgent and non-urgent jobs. Higher
the value of $p$, higher is the dominance of urgent jobs over non-urgent jobs. GPSU algorithm tries to mimic UDSU, NUBSU
and FSU algorithm based on probability parameter $p$.\\
\hspace*{5 mm}For $p = 1$, GPSU algorithm mimics the nature of UDSU algorithm. The algorithm will completely ignore the non urgent
jobs in the presence of urgent jobs and will lead to their starvation in the presence of frequent arrival of urgent jobs. For higher
values of $p < 1$ , the algorithm will pick urgent jobs with very high probability, however at the same time, the algorithm 
will not create starvation for non urgent jobs since the slowed down non urgent jobs will be picked up by their 
priority function (low arrival time resulting in low priority function value) with probability ($1$-$p$) thereby mimicking
NUBSU algorithm.
For $p$ proportional to the fraction of total jobs requesting for speed up (urgent jobs), the algorithm will behave as 
FSU algorithm since we are giving equal chance to both urgent as well as non-urgent jobs.\\
\hspace*{5 mm}The algorithm can be implemented by using two separate \textit{priority queues} one for the urgent jobs (based on
$P_{urgent}(j)$ function value as per eq. \ref{Purgenteq} ) and the other one for non-urgent
jobs (based on P function value defined in eq. \ref{Pfunction}). The time complexity of this algorithm is thus $O(log($ $\left|{J_{urgent}}(t)\right|)$ $+$ $log($ $\left|{J_{nonurgent}}(t)\right|)$ $)$
to pick up the next job at time $t$ and is computationally efficient. It is important to note that GPSU algorithm can also be used for Speed up with no 
constraints scenario, with only one minor change
that instead of using $P_{urgent}$, use earlier defined $P$ priority function defined in eq. \ref{Pfunction} in the GPSU algorithm. 
The NUBSU algorithm had high time complexity and thus,
GPSU algorithm can be used to mimic its nature by setting the probability parameter $p$. \\
\hspace*{5 mm}Value of $p$ to be used in GPSU algorithm depends on the required preference/dominance of urgent over non-urgent jobs.
If jobs requesting for speed up are extremely critical and urgent, then $p$ should be $1$ or very close to it. However, if
jobs requesting for speed up are not much critical and high slow down for any of the job might have severe consequences then in that case
value of $p$
can be set somewhere between $0.6$ - $0.9$. Value of probability parameter is thus dependent on the requirement of the environment
based on the amount of slow down that can be tolerated and the amount of speed up which is intended/desired by the system.
\section{Experiments and Results}
\hspace*{5 mm}We used a discrete event based simulation tool for the purpose of comparison analysis and evaluation of our speed up
algorithms. We assumed that jobs are arriving in Poisson distribution. The total number of jobs we considered for the
experiments were 10 thousand. The service time of jobs were assumed to be in exponential distribution. We conducted 
experiments for three different values of system load $\rho$ ($\lambda / \mu)$ equal to 1.1, 1.3 and 1.5. Since the nature of
results obtained were found similar for different $\rho$ values, therefore we present here the results only for $\rho$ value
equal to $1.1$. We labeled some of the arriving jobs among total jobs as urgent and conducted experiments by varying proportion of urgent
jobs (jobs requesting for speed up) from $0\%$ to $100\%$. Value of $\mu$ used is $0.04$ and $\lambda$ varies depending
on the value of $\rho$.\\
\subsection{Speed Up with no constraints}
\subsubsection{Percentage of urgent jobs that achieved speed up}
\begin{table}
\centering
\tbl{$\rho = 1.1$: Percentage of urgent jobs that achieved speed up.\label{case1:overallspeedup-1}}{
\setlength{\tabcolsep}{8pt}
\begin{tabular}{|c|c|c|c|c|c|l|} \hline
Urgent jobs&MPF 	&MPF-SD			&MinPF			&UDSU			&NUBSU			&FSU\\ \hline
0\%&-			&-			&-			&-			&-			&- \\ \hline
10\%&\textbf{97\%}	&94\%			&\textbf{97\%}		&\textbf{97\%}		&94\%			&84\% \\ \hline
20\%&\textbf{95\%}	&91.5\%			&\textbf{95\%}		&\textbf{95.5\%}	&91.5\%			&84\% \\ \hline
30\%&\textbf{94.67\%}	&93.3\%			&\textbf{94.33\%}	&\textbf{95\%}		&90\%			&84.33\% \\  \hline
40\%&93.5\%		&93.5\%			&\textbf{94.25\%}	&\textbf{94.5\%}	&89.75\%		&85.25\% \\ \hline
50\%&94.2\%		&94.6\%			&\textbf{95.2\%}	&\textbf{95\%}		&90.6\%			&86.6\% \\ \hline
60\%&93.67\%		&92.83\%		&\textbf{93.83\%}	&\textbf{94.5\%}	&90.5\%			&86.33\% \\ \hline
70\%&92.71\%		&92\%			&\textbf{93.86\%}	&\textbf{94\%}		&89.75\%		&87\% \\ \hline
80\%&91.62\%		&90.62\%		&\textbf{93\%}		&\textbf{93\%}		&90.88\%		&87.25\% \\ \hline
90\%&89.78\%		&89.67\%		&\textbf{90.89\%}	&\textbf{90.56\%}	&90.56\%		&87.56\% \\ \hline
100\%&-			&-			&-			&87.3\%			&87.3\%		 	&87.3\%\\ \hline
\end{tabular}}
\end{table}
The Table \ref{case1:overallspeedup-1} gives the percentage of urgent jobs (jobs requesting for speed up) that achieved speed up for different speed up algorithms based on the proportion of urgent jobs for system load 
$\rho = 1.1$. UDSU algorithm is successful in achieving near about similar (sometimes even higher) amount of speed up as compared to existing
speed up algorithms MPF and MinPF. The UDSU algorithm outperforms MPF-SD in terms of achieved speed up for majority of urgent job
proportion. It is interesting to note that the performance of NUBSU algorithm is lower than UDSU algorithm and other existing speed up algorithms in terms
of achieved speed up validating our idea that the acceleration of opportunistically forwardable job does not always ensure 
that every other existing urgent job present in the queue will achieve speed up and thus, such acceleration may even slow down the urgent jobs. But still,
NUBSU algorithm speeds up large number of urgent jobs. The percentage of speeded up urgent jobs is lowest for FSU algorithm since it is fair to all and is not biased towards urgent jobs.
When all the jobs present in the queue are requesting for speed up ($100\%$ case), then all the three speed up (UDSU, NUBSU
and FSU) algorithm behaves as same. As the percentage of urgent jobs requesting for speed up increases, achieved speed up
decreases.
\subsubsection{Percentage of slowed down non urgent jobs}
Table \ref{case1:overallslowdown-1} shows that our implicit
speed up algorithms (UDSU, NUBSU and FSU) performs better than MPF, MPF-SD and MinPF algorithms in terms of slow down caused
to the other non-urgent jobs present in the queue. The reason is that MPF, MPF-SD and MinPF algorithms do
not consider about the slow down of other non-urgent jobs and greedily aim at achieving high speed up for urgent jobs
using location table and swapping of jobs. However, in our speed up algorithms,the non-urgent jobs are served using the 
priority function which takes into consideration the arrival time (to avoid high wait time) as well as duration
time (to improve the mean wait time) which avoids high slow down for non urgent jobs. Note that NUBSU algorithm performs better than UDSU algorithm in terms of slow
down since NUBSU algorithm whenever it finds some non-urgent job to be opportunistically forwardable, it accelerates 
it to the server resulting in less percentage of slow downs for non-urgent jobs as compared to UDSU algorithm. It is also
observed that as the proportion of urgent jobs increases, the slow down caused to non-urgent jobs also increases. FSU algorithm
provides least amount of slow down to non urgent jobs since FSU is fair to all and is not biased towards any
specific job.
\begin{table}
\centering
\tbl{$\rho = 1.1$: Percentage of slowed down non-urgent jobs.\label{case1:overallslowdown-1}}{
\setlength{\tabcolsep}{8pt}
\begin{tabular}{|c|c|c|c|c|c|l|} \hline
Urgent jobs&MPF 	&MPF-SD			&MinPF			&UDSU			&NUBSU			&FSU\\ \hline
0\%&-			&-			&-			&\textbf{8.8\%}		&\textbf{8.8\%}		&\textbf{8.8\%} \\ \hline
10\%&65.78\%		&\textbf{8.56\%}	&89.11\%		&\textbf{10.89\%}	&\textbf{9.56\%}	&\textbf{8.44\%} \\ \hline
20\%&67\%		&15.25\%		&94.62\%		&\textbf{13.38\%}	&\textbf{10.25\%}	&\textbf{8.12\%} \\ \hline
30\%&71.86\%		&24.57\%		&95.67\%		&\textbf{16.29\%}	&\textbf{13.14\%} 	&\textbf{7.71\%} \\  \hline
40\%&68.67\%		&31\%			&95.83\%		&\textbf{22\%}		&\textbf{16.5\%}	&\textbf{7.83\%} \\ \hline
50\%&75\%		&39.8\%			&96.2\%			&\textbf{27\%}		&\textbf{22.8\%}	&\textbf{8.4\%} \\ \hline
60\%&80\%		&51.5\%			&96.75\%		&\textbf{35.25\%}	&\textbf{35\%}		&\textbf{8\%} \\ \hline
70\%&82\%		&49\%			&96\%			&\textbf{46\%}		&\textbf{41.67\%}	&\textbf{8\%} \\ \hline
80\%&84.5\%		&60\%			&96.5\%			&\textbf{60}\%		&\textbf{59\%}		&\textbf{9\%} \\ \hline
90\%&88\%		&82\%			&97.4\%			&86\%			&90\%			&\textbf{12\%} \\ \hline
100\%&-			&-			&-			&-			&-		 	&-\\ \hline
\end{tabular}}
\end{table}
\subsubsection{Mean Wait Time}
Table \ref{case1:meanwaittime-1} shows the overall mean wait time
for all the jobs present in the queue for different urgent job proportion. The mean wait time for FSU algorithm is lowest owing to its unbiased
nature and it remains unaffected with respect to the proportion of urgent jobs since
the algorithm is independent of it. Our speed up algorithms outperforms existing position based speed up algorithms (MPF, MinPF and MPF-SD)
in terms of overall mean wait time owing to the design of the priority function which prefers job with shorter duration time (resulting in
low mean wait time).
\begin{table}
\centering
\tbl{$\rho = 1.1$: Mean Wait Time (in time units).\label{case1:meanwaittime-1}}{
\setlength{\tabcolsep}{8pt}
\begin{tabular}{|c|c|c|c|c|c|l|} \hline
Urgent jobs&MPF 	&MPF-SD			&MinPF			&UDSU			&NUBSU			&FSU\\ \hline
0\%&-			&-			&-			&\textbf{473}		&\textbf{473}		&\textbf{473} \\ \hline
10\%&1742		&1575			&1713			&\textbf{495}		&\textbf{535}		&\textbf{473} \\ \hline
20\%&1738		&1459			&1711			&\textbf{536}		&\textbf{602}		&\textbf{473} \\ \hline
30\%&1715		&1372			&1694			&\textbf{558}		&\textbf{668}	 	&\textbf{473} \\  \hline
40\%&1709		&1288			&1692			&\textbf{584}		&\textbf{759}		&\textbf{473} \\ \hline
50\%&1747		&1164			&1675			&\textbf{633}		&\textbf{779}		&\textbf{473} \\ \hline
60\%&1705		&1137			&1669			&\textbf{738}		&\textbf{848}		&\textbf{473} \\ \hline
70\%&1672		&1069			&1625			&\textbf{823}		&\textbf{876}		&\textbf{473} \\ \hline
80\%&1670		&1073			&1591			&\textbf{925}		&\textbf{897}		&\textbf{473} \\ \hline
90\%&1562		&1259			&1548			&\textbf{1111}		&\textbf{1111}		&\textbf{473} \\ \hline
100\%&-			&-			&-			&\textbf{473}		&\textbf{473}		&\textbf{473}\\ \hline
\end{tabular}}
\end{table}

\subsubsection{95 percentile metric for wait time}
\begin{table}
\centering
\tbl{$\rho = 1.1$: 95 percentile metric/maximum for wait time (in time units).\label{case1:95percentile-1}}{
\setlength{\tabcolsep}{3pt}
\begin{tabular}{|c|c|c|c|c|c|l|} \hline
Urgent jobs&MPF 		&MPF-SD			&MinPF			&UDSU			&NUBSU			&FSU\\ \hline
0\%	   &-			&-			&-			&\textbf{3269 / 14002}	&\textbf{3269 / 14002}	&\textbf{3269 / 14002} \\ \hline
10\%&4018 / 7817		&3857 / 10160		&3911 / 4125		&\textbf{3421 / 14218}	&\textbf{3449 / 14008}	&\textbf{3269 / 14002} \\ \hline
20\%&5085 / 11657		&3985 / 13053		&4136 / 4263		&\textbf{3661 / 14533}	&\textbf{3307 / 14394}	&\textbf{3269 / 14002} \\ \hline
30\%&6119 / 15324		&4505 / 12132		&4491 / 4644		&\textbf{3875 / 14825}	&\textbf{3254 / 14498} 	&\textbf{3269 / 14002} \\  \hline
40\%&6770 / 15854		&4929 / 14302		&5423 / 6011		&\textbf{3658 / 16840}	&\textbf{3503 / 14829}	&\textbf{3269 / 14002} \\ \hline
50\%&8257 / 17293		&5303 / 17525		&6049 / 6959		&\textbf{3822 / 17860}	&\textbf{3660 / 16049}	&\textbf{3269 / 14002} \\ \hline
60\%&8970 / 14967		&5495 / 19537		&7046 / 7695		&\textbf{4873 / 16533}	&\textbf{4002 / 16452}	&\textbf{3269 / 14002} \\ \hline
70\%&9751 / 21812		&6020 / 19069		&8321 / 9905		&\textbf{5707 / 17281}	&\textbf{4106 / 16798}	&\textbf{3269 / 14002} \\ \hline
80\%&11355 / 22886		&6526 / 21901		&9638 / 12983		&\textbf{6485 / 17795}	&\textbf{4750 / 17795}	&\textbf{3269 / 14002} \\ \hline
90\%&9615 / 21751		&7308 / 19959		&9875 / 18488		&10041 / 20491		&9872 / 20491		&\textbf{3269 / 14002} \\ \hline
100\%&-			&-			&-			&\textbf{3269 / 14002}		&\textbf{3269 / 14002}	&\textbf{3269 / 14002}\\ \hline
\end{tabular}}
\end{table}
Table \ref{case1:95percentile-1} indicates
the 95 percentile metric of wait time for different speed up algorithms. Experimental results show that MinPF algorithm
provides the least maximum wait time among all the existing algorithms. The reason is MinPF algorithm swap jobs
with minimum distance. Our implicit speed up algorithms (UDSU, NUBSU and FSU) outperforms the existing speed up algorithms 
in terms of 95 percentile metric owing to the design of the priority function. 
\subsubsection{DNUJI for NUBSU algorithm}
\begin{figure}
\centerline{\includegraphics[width=120mm, height = 70mm]{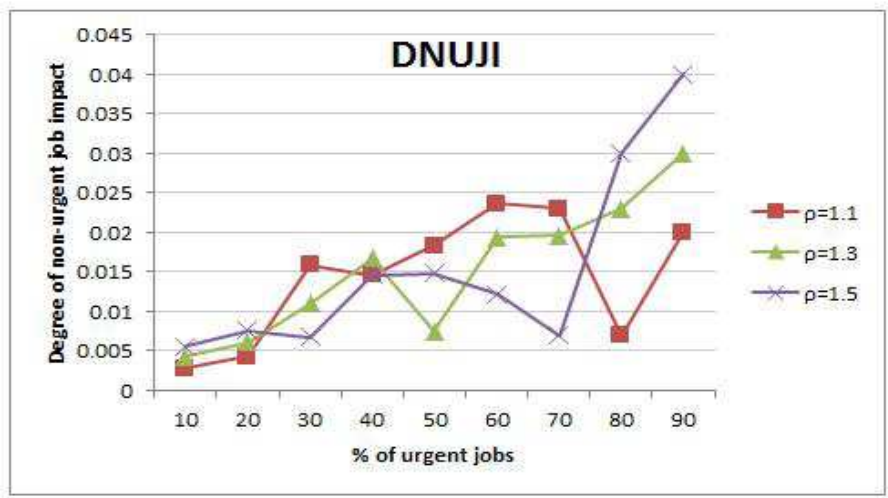}}
\caption{Degree of non-urgent job impact (DNUJI) for NUBSU algorithm for different values of system load.}
\label{DNUJI}
\end{figure}
Figure \ref{DNUJI} shows the plot of DNUJI for NUBSU algorithm based on proportion of urgent jobs for different values
of system load $\rho$. The non-zero value of DNUJI validates our idea that acceleration of opportunistically forwardable
job do not always ensure that every other urgent job would achieve speed up. But, experimentally it still happens even 
less than 5 times out of 100.
\subsection{Fine Grained Selective Speed Up}
\hspace*{5 mm}In order to consider Fine Grained Selective Speed Up scenario, among all the urgent jobs (requesting for speed up)
we assigned each of them random percentage value $pj \in (0, 100]$ corresponding to which their RSU will be calculated.
Note that existing MPF, MPF-SD and MinPF algorithms are not applicable for cases like Fine Grained Selective Speed Up where each job $j$
requests for some specific $p_j\%$ of max speed up that it can achieve. Therefore, we present here the results for our
GPSU algorithm for different values of probability parameter $p$.
\subsubsection{Percentage of successfully speeded up urgent jobs}
Table \ref{case2:speedup-1} shows the percentage of successfully
speeded up urgent jobs (jobs for which achieved speed up $>=$ Requested speed up ) for GPSU algorithm for different values of
probability parameter $p$. For $p=1$, maximum successful speed ups were obtained. The percentage of successful speed ups decreases as probability
parameter $p$ decreases, since the tendency to favor urgent jobs decreases with decrease in $p$. It is also experimentally
observed, that as the proportion of urgent jobs requesting for speed up increases, percentage of successful speed ups decreases.
\begin{table}
\centering
\tbl{$\rho = 1.1$: Percentage of successfully speeded up urgent jobs for GPSU Algorithm.\label{case2:speedup-1}}{
\setlength{\tabcolsep}{8pt}
\begin{tabular}{|c|c|c|c|l|} \hline
Urgent jobs 	&p=1			&p=0.9			&p=0.8			&p=0.7 \\ \hline
0\% 		&-			&-			&-			&- \\ \hline
10\%		&\textbf{94\%}		&\textbf{94\%}		&94\%			&92\% \\ \hline
20\% 		&\textbf{92\%}		&\textbf{91.5\%}	&91\%			&89\% \\ \hline
30\%		&\textbf{92\%}		&\textbf{91.33\%}	&91\%			&90\% \\ \hline
40\%		&\textbf{91.75\%}	&\textbf{90.5\%}	&89.25\%		&88.25\% \\ \hline
50\%		&\textbf{92.6\%}	&\textbf{91.8\%}	&90.8\%			&82.8\% \\ \hline
60\%		&\textbf{90.17\%}	&\textbf{87.83\%}	&85.17\%		&76\% \\ \hline
70\%		&\textbf{87.57\%}	&\textbf{85.57\%}	&82.43\%		&60.43\% \\ \hline
80\%		&\textbf{81.5\%}	&\textbf{79.12\%}	&36.75\%		&2.2\% \\ \hline
90\%		&\textbf{57.11\%}	&\textbf{5\%}		&3.3\%			&2.11\% \\ \hline
100\% 		&\textbf{2.2\%}		&\textbf{2.2\%}		&2.2\%			&2.2\% \\ \hline
\end{tabular}}
\end{table}
\subsubsection{Percentage of slowed down non urgent jobs}
Table \ref{case2:slowdown-1} shows the percentage of slowed down
non-urgent jobs for GPSU Algorithm for different values of probability parameter $p$. Higher the probability parameter,
higher is the percentage of slowed down non-urgent jobs since as $p$ increases, tendency to serve non-urgent jobs decreases which 
leads to their slow down. Thus, the percentage of slowed down jobs which are not requesting for speed up is least for $p=0.7$.
\begin{table}
\centering
\tbl{$\rho = 1.1$: Percentage of slowed down non-urgent jobs for GPSU Algorithm.\label{case2:slowdown-1}}{
\setlength{\tabcolsep}{8pt}
\begin{tabular}{|c|c|c|c|l|} \hline
Urgent jobs 	&p=1			&p=0.9			&p=0.8			&p=0.7 \\ \hline
0\% 		&8.8\%			&8.8\%			&\textbf{8.8\%}		&\textbf{8.8\%} \\ \hline
10\%		&10.89\%		&10.89\%		&\textbf{10.78\%}	&\textbf{10.78\%} \\ \hline
20\% 		&13.38\%		&13.38\%		&\textbf{13.25\%}	&\textbf{13\%} \\ \hline
30\%		&16.29\%		&15.86\%		&\textbf{15.43\%}	&\textbf{15.43\%} \\ \hline
40\%		&22\%			&21.33\%		&\textbf{20.17\%}	&\textbf{19.67\%} \\ \hline
50\%		&27\%			&25.8\%			&\textbf{24.4\%}	&\textbf{23.4\%} \\ \hline
60\%		&35.25\%		&35\%			&\textbf{32.25\%}	&\textbf{28\%} \\ \hline
70\%		&46\%			&41.33\%		&\textbf{37.33\%}	&\textbf{28.67\%} \\ \hline
80\%		&64\%			&55\%			&\textbf{32\%}		&\textbf{5.5\%} \\ \hline
90\%		&90\%			&46\%			&\textbf{9\%}		&\textbf{2\%} \\ \hline
100\% 		&-			&-			&-			&- \\ \hline
\end{tabular}}
\end{table}

\begin{table}
\centering
\tbl{$\rho = 1.1$: Overall mean wait time for all the jobs for GPSU Algorithm.\label{case2:meanwaittime-1}}{
\setlength{\tabcolsep}{8pt}
\begin{tabular}{|c|c|c|c|l|} \hline
Urgent jobs 	&p=1			&p=0.9			&p=0.8			&p=0.7 \\ \hline
0\% 		&473			&473			&473			&473 \\ \hline
10\%		&495			&495			&495			&495 \\ \hline
20\% 		&537			&536			&530			&537 \\ \hline
30\%		&560			&559			&560			&559 \\ \hline
40\%		&589			&588			&591			&587 \\ \hline
50\%		&645			&642			&644			&644 \\ \hline
60\%		&760			&749			&752			&760 \\ \hline
70\%		&868			&850			&847			&947 \\ \hline
80\%		&1039			&984			&1129			&1711 \\ \hline
90\%		&1532			&1441			&1728			&1751 \\ \hline
100\% 		&1717			&1717			&1717			&1717 \\ \hline
\end{tabular}}
\end{table}

\subsubsection{Mean Wait time}
Table \ref{case2:meanwaittime-1} shows the overall mean wait time (in time units) 
for all the jobs for GPSU Algorithm for different probability values.
\subsection{Experimentation on process logs dataset}
Apart from conducting experiments on M/M/1 queueing system,
we also conducted experiments on real process logs obtained using \textit{Process Accounting} utility in linux. For the purpose
of gathering logs, we executed different types of user programs (dynamic programming, recursion, modular exponentiation, binary
search , adhoc etc.)
randomnly in poisson fashion using shell script on randomnly generated datasets at run time. The RAM of the system is
3GB and processor used is 2.4 GHz. Using pacct utility in linux, 
we collected process logs for about 10K executed processes which involves both user as well as kernel processes.
The minimum and maximum cpu burst time obtained in the logs were 1 and 185 units respectively. Some of the sample log records
are shown in table \ref{pacct},
\begin{table}
\centering
\tbl{Sample process log records obtained using pacct utility in linux.\label{pacct}}{
\setlength{\tabcolsep}{3pt}
\begin{tabular}{|c|c|c|c|c|c|c|c|c|c|c|l|} \hline
command &version  &utime &systime &etime &uid  &gid  &mem      &char    &pid    &ppid   &finish time \\ \hline
cc1plus &v3       &17.00 &1.00    &19.00 &1000 &1000 &31632.00 &0.00         &27379  &27378  &Sun Dec 15 16:47:50 2013 \\ \hline
a.out   &v3       &16.00 &0.00    &16.00 &1000 &1000 &7988.00  &0.00         &27389  &27324  &Sun Dec 15 16:47:51 2013 \\ \hline
ld      &v3       &5.00  &0.00    &6.00  &1000 &1000 &10512.00 &0.00         &27554  &27553  &Sun Dec 15 16:51:02 2013  \\ \hline
\end{tabular}}
\end{table}
where command - command name, version - version of acct file, utime - user time, systime - system time, etime - elapsed time, uid - user id,
gid - group id, mem - memory usage, char - number of characters transferred on input/output, pid - process id, 
ppid - parent pid, finish time - finish time of process.
The nature of results obtained are similar as compared to M/M/1 system and are presented here.
\begin{table}
\centering
\tbl{Process logs dataset: Percentage of urgent jobs that achieved speed up for Speed up with no constrains scenario.\label{case1processlogsdatset:speedup}}{
\setlength{\tabcolsep}{8pt}
\begin{tabular}{|c|c|c|c|c|c|l|} \hline
Urgent jobs&MPF 	&MPF-SD			&MinPF			&UDSU			&NUBSU			&FSU\\ \hline
0\%&-			&-			&-			&-			&-			&- \\ \hline
10\%&\textbf{98.99\%}	&97.98\%		&\textbf{99.8\%}	&\textbf{98.99\%}	&97.98\%		&82.83\% \\ \hline
20\%&\textbf{98.99\%}	&98.48\%		&\textbf{99.49\%}	&\textbf{98.99\%}	&95.96\%		&79.29\% \\ \hline
30\%&98.32\%		&\textbf{98.99\%}	&\textbf{99.33\%}	&\textbf{99.2\%}	&94.28\%		&79.12\% \\  \hline
40\%&99.14\%		&\textbf{99.75\%}	&\textbf{99.49\%}	&\textbf{99.24\%}	&95.71\%		&85.75\% \\ \hline
50\%&98.79\%		&\textbf{99.39\%}	&98.79\%		&\textbf{99.39\%}	&96.16\%		&80.61\% \\ \hline
60\%&98.65\%		&\textbf{99.16\%}	&98.65\%		&\textbf{99.1\%}	&97.47\%		&80.64\% \\ \hline
70\%&98.56\%		&\textbf{98.99\%}	&98.56\%		&\textbf{98.67\%}	&96.97\%		&80.95\% \\ \hline
80\%&\textbf{98.48\%}	&\textbf{99.12\%}	&\textbf{98.48\%}	&96.09\%		&95.96\%		&81.44\% \\ \hline
90\%&\textbf{98.32\%}	&\textbf{98.32\%}	&\textbf{98.2\%}	&91.69\%		&90.69\%		&81.14\% \\ \hline
100\%&-			&-			&-			&81.33\%		&81.33\%	 	&81.33\%\\ \hline
\end{tabular}}
\end{table}

\begin{table}
\centering
\tbl{Process logs dataset: Percentage of slowed down non-urgent jobs for speed up with no constraints scenario.\label{case1processlogsdatset:slowdown}}{
\setlength{\tabcolsep}{8pt}
\begin{tabular}{|c|c|c|c|c|c|l|} \hline
Urgent jobs&MPF 	&MPF-SD			&MinPF			&UDSU			&NUBSU			&FSU\\ \hline
0\%&-			&-			&-			&\textbf{18.16\%}	&\textbf{18.16\%}	&\textbf{18.16\%} \\ \hline
10\%&84.75\%		&\textbf{9.08\%}	&97.98\%		&20.63\%		&20.29\%		&\textbf{18.27\%} \\ \hline
20\%&81.2\%		&\textbf{17.78\%}	&98.87\%		&22.45\%		&20.3\%			&\textbf{17.65\%} \\ \hline
30\%&79.25\%		&25.94\%		&99.33\%		&\textbf{25.5\%}	&\textbf{22.48\%} 	&\textbf{17.15\%} \\  \hline
40\%&68.57\%		&34.12\%		&99.49\%		&\textbf{33.61\%}	&\textbf{23.87\%}	&\textbf{12.67\%} \\ \hline
50\%&71.77\%		&45.16\%		&99.6\%			&\textbf{40.52\%}	&\textbf{29.64\%}	&\textbf{17.14\%} \\ \hline
60\%&70.78\%		&54.66\%		&98.79\%		&\textbf{52.53\%}	&\textbf{51.21\%}	&\textbf{17.38\%} \\ \hline
70\%&93.96\%		&79.8\%			&99.7\%			&\textbf{78.52\%}	&\textbf{72.48\%}	&\textbf{17.45\%} \\ \hline
80\%&80.4\%		&86.88\%		&99.73\%		&86.93			&\textbf{83.92\%}	&\textbf{18.59\%} \\ \hline
90\%&93\%		&87\%			&99.8\%			&93\%			&93\%			&\textbf{17\%} \\ \hline
100\%&-			&-			&-			&-			&-		 	&-\\ \hline
\end{tabular}}
\end{table}

\begin{table}
\centering
\tbl{Process logs dataset: Mean Wait Time for Speed Up with no constraints scenario (in time units).\label{case1processlogsdatset:meanwaittime}}{
\setlength{\tabcolsep}{8pt}
\begin{tabular}{|c|c|c|c|c|c|l|} \hline
Urgent jobs&MPF 	&MPF-SD			&MinPF			&UDSU			&NUBSU			&FSU\\ \hline
0\%&-			&-			&-			&\textbf{1122}		&\textbf{1122}		&\textbf{1122} \\ \hline
10\%&3507		&3154			&3489			&\textbf{1185}		&\textbf{1170}		&\textbf{1122} \\ \hline
20\%&3337		&2765			&3528			&\textbf{1334}		&\textbf{1291}		&\textbf{1122} \\ \hline
30\%&3573		&2752			&3606			&\textbf{1516}		&\textbf{1425}	 	&\textbf{1122} \\  \hline
40\%&3484		&2556			&3492			&\textbf{1599}		&\textbf{1529}		&\textbf{1122} \\ \hline
50\%&3420		&2518			&3485			&\textbf{1735}		&\textbf{1631}		&\textbf{1122} \\ \hline
60\%&3484		&2672			&3573			&\textbf{2484}		&\textbf{2235}		&\textbf{1122} \\ \hline
70\%&3438		&2716			&3466			&\textbf{2416}		&\textbf{2315}		&\textbf{1122} \\ \hline
80\%&3215		&2870			&3261			&\textbf{2186}		&\textbf{2010}		&\textbf{1122} \\ \hline
90\%&3319		&3100			&3335			&\textbf{1763}		&\textbf{1763}		&\textbf{1122} \\ \hline
100\%&-			&-			&-			&\textbf{1122}		&\textbf{1122}		&\textbf{1122}\\ \hline
\end{tabular}}
\end{table}

\begin{table}
\centering
\tbl{Process logs dataset: Percentage of successfully speeded up urgent jobs for GPSU Algorithm.\label{case2processlogsdatset:speedup}}{
\setlength{\tabcolsep}{8pt}
\begin{tabular}{|c|c|c|c|l|} \hline
Urgent jobs 	&p=1			&p=0.9			&p=0.8			&p=0.7 \\ \hline
0\% 		&-			&-			&-			&- \\ \hline
10\%		&\textbf{97.98\%}	&\textbf{97.98\%}	&96.97\%		&96.97\% \\ \hline
20\% 		&\textbf{97.98\%}	&\textbf{97.91\%}	&96.97\%		&96.46\% \\ \hline
30\%		&\textbf{97.93\%}	&\textbf{97.64\%}	&97.63\%		&96.64\% \\ \hline
40\%		&\textbf{97.22\%}	&\textbf{97.22\%}	&96.97\%		&95.96\% \\ \hline
50\%		&\textbf{95.96\%}	&\textbf{93.74\%}	&93.15\%		&92.73\% \\ \hline
60\%		&\textbf{93.43\%}	&\textbf{93.77\%}	&93.1\%			&92.42\% \\ \hline
70\%		&\textbf{92.06\%}	&\textbf{91.92\%}	&90.62\%		&1.01\% \\ \hline
80\%		&\textbf{56.68\%}	&\textbf{16.54\%}	&0.51\%			&0.38\% \\ \hline
90\%		&\textbf{1.97\%}	&\textbf{1.01\%}	&0.45\%			&0.34\% \\ \hline
100\% 		&\textbf{0.3\%}		&\textbf{0.3\%}		&0.3\%			&0.3\% \\ \hline
\end{tabular}}
\end{table}

\begin{table}
\centering
\tbl{Process logs dataset: Percentage of slowed down non-urgent jobs for GPSU Algorithm.\label{case2processlogsdatset:slowdown}}{
\setlength{\tabcolsep}{8pt}
\begin{tabular}{|c|c|c|c|l|} \hline
Urgent jobs 	&p=1			&p=0.9			&p=0.8			&p=0.7 \\ \hline
0\% 		&18.16\%		&18.16\%		&\textbf{18.16\%}	&\textbf{18.16\%} \\ \hline
10\%		&20.63\%		&20.63\%		&\textbf{20.21\%}	&\textbf{20.21\%} \\ \hline
20\% 		&22.45\%		&22.45\%		&\textbf{22.32\%}	&\textbf{22.32\%} \\ \hline
30\%		&25.5\%			&25.4\%			&\textbf{25.36\%}	&\textbf{25.21\%} \\ \hline
40\%		&33.61\%		&33.6\%			&\textbf{33.51\%}	&\textbf{33.45\%} \\ \hline
50\%		&40.52\%		&41.13\%		&\textbf{40.32\%}	&\textbf{40.12\%} \\ \hline
60\%		&70.53\%		&64.48\%		&\textbf{48.61\%}	&\textbf{43.83\%} \\ \hline
70\%		&78.52\%		&72.82\%		&\textbf{48.66\%}	&\textbf{28.52\%} \\ \hline
80\%		&86.93\%		&51.79\%		&\textbf{17.09\%}	&\textbf{1.01\%} \\ \hline
90\%		&93\%			&43\%			&\textbf{2\%}		&\textbf{0.2\%} \\ \hline
100\% 		&-			&-			&-			&- \\ \hline
\end{tabular}}
\end{table}

\begin{table}
\centering
\tbl{Process logs dataset: Overall mean wait time for all the jobs for GPSU Algorithm.\label{case2processlogsdatset:meanwaittime}}{
\setlength{\tabcolsep}{8pt}
\begin{tabular}{|c|c|c|c|l|} \hline
Urgent jobs 	&p=1			&p=0.9			&p=0.8			&p=0.7 \\ \hline
0\% 		&1122			&1122			&1122			&1122 \\ \hline
10\%		&1186			&1186			&1185			&1186 \\ \hline
20\% 		&1335			&1335			&1334			&1334 \\ \hline
30\%		&1518			&1517			&1517			&1518 \\ \hline
40\%		&1603			&1596			&1595			&1587 \\ \hline
50\%		&1750			&1730			&1720			&1735 \\ \hline
60\%		&2571			&2340			&2138			&1980 \\ \hline
70\%		&2806			&2596			&2377			&2453 \\ \hline
80\%		&3003			&2596			&2847			&3237 \\ \hline
90\%		&3196			&2924			&3254			&3293 \\ \hline
100\% 		&3297			&3297			&3297			&3297 \\ \hline
\end{tabular}}
\end{table}

\subsection{Discussion}
The authors in \cite{kafeza} and \cite{wecwis} used only achieved ratio metric (percentage of urgent jobs that achieve 
speed up) for evaluating the effectiveness of their speed up algorithms. In this research, we along with achieved ratio metric
also consider the amount of slow down for rest of the jobs as a result of speed up and overall mean wait time. The experimental results show that our proposed speed up algorithm UDSU is able to achieve near about similar amount of speed
up as compared to the existing speed up algorithms MPF, MPF-SD and MinPF. All the three speed up algorithms UDSU, NUBSU
and FSU outperforms the existing speed up algorithms in terms of slow down caused to the other non-urgent jobs. It is because
existing speed up algorithms do not provide any remedies for delayed jobs and greedily aim at achieving high speed up for urgent jobs.
However, our implicit speed up priority function takes into consideration the arrival time (while speeding up urgent jobs)
which avoids high slow down for non-urgent jobs. The overall mean wait time is improved dramatically by our speed up algorithms
owing to the design of priority function as compared to the existing speed up algorithms which are purely positional. Further, our algorithms are implicit
in nature where the notion of acceleration is incorporated in the priority function which leads to a time efficient solution
unlike existing algorithms where there is an overhead of swapping of jobs and \textit{location table}. We further presented the
GPSU algorithm for \textit{Fine Grained Selective Speed Up} scenario which has not been yet addressed and presented the
results for the same.\\
\hspace*{5 mm}We used our experimental methodology to be simulation. The experiments performed involved changing various parameters (
system load, jobs requesting for speed up, RSU etc.) and the results shown are the statistical averages of multiple runs. We conducted our experiments assuming a M/M/1 queuing 
system which is widely used queueing system for the analysis of real life situations and is a good approximation for large number of 
queueing systems \cite{booktheory}. The poisson arrival distribution of jobs assumed in experiments is a very good approximation for job arrival in real systems.
Regarding exponential distribution there is an argument from information theory which states that exponential distribution provides
the least information or highest entropy, and is therefore a reasonable assumption for service time when no other data is available
\cite{booktheory}. Thus, our assumption of M/M/1 system for experimental analysis is valid to a greater extent and our
results are significant. We also achieved similar nature of results using real process logs dataset. Further, we also have considered real trace data for experimental analysis presented in later chapters of this thesis. 

\section{Applications: Web Scheduling}
\hspace*{5 mm}We apply the idea of implicit speed up for the purpose of scheduling of web requests arriving at 
web server. We show how our proposed priority function (defined in eq.\ref{Pfunction}) for implicit speed up with little modification, can be applied to
web scheduling. We show the effectiveness of our algorithms using simulation model on a trace driven workload.
\subsection{Introduction}
\hspace*{5 mm}Current demand on busy web servers requires them to
serve up to thousands of clients simultaneously. In such a
case, the response time suffered by the client is one of the
most important factor that determines the web server performance, where the response time is defined as the time duration 
between the time client makes the request until the
time the client receives the last byte of the file requested.
The servers cannot afford large delay in response time for
users because it might result in rejection of request either
because of server timeout or due to user abort. The slow response times and difficult navigation are the most common
complaints of Internet Users \cite{website}. The scheduling strategies
used by a web server play a very crucial role in determining
the response time of users. Now, the question that arises
is how to improve the response time of the user but at the
same time ensuring no starvation by simply changing the
order of servicing of requests arriving at web server.\\
\hspace*{5 mm}The traditional scheduling strategy used in web server
is Processor-Sharing (PS) scheduling in which each of the $n$
incoming requests gets same amount $1/n$ of the CPU time.
PS scheduling is unbiased as it gives equal proportion of
time to all.\\
\hspace*{5 mm}It is well known from queuing theory that Shortest Remaining Processing Time (SRPT) scheduling is an optimal scheduling
algorithm for minimizing the mean response time \cite{proof}. SRPT based scheduling policies uses size of requested file to implement SRPT.
Various SRPT based scheduling policies \cite{crovella}, \cite{balter}, \cite{2006srpt}, \cite{2007srpt} and \cite{srrt} have been proposed for the scheduling of HTTP requests arriving at web server. Still 
these algorithms are not used in practice. There are several reasons for this. One reason is that SRPT causes starvation for large
size requests (because of its tendency to favor small size requests) in the presence of continuous arrival of small sized requests.
Also, SRPT relies completely on size of requested file to determine the mean response time which is not enough because it does not
take into consideration the user server interaction parameters present in Internet like throughput of the user connection. Further,
SRPT based policies cannot be applied for dynamic HTTP requests where file size is not known in advance.
\subsection{Related Work}
\hspace*{5 mm}L. Cherkasova introduced the concept of Alpha-Scheduling
to improve the response time for web applications \cite{alpha}. This
strategy lies between FIFO (first-in-first-out) and SRPT.
The measure of balance between FIFO and SRPT was decided by the parameter alpha.The major drawback of this strategy is that
they have not taken into consideration the user-server interaction parameters over Internet like network bandwidth,
which may play a crucial role in influencing the response
time of the user.\\
\hspace*{5 mm}It is well known from queuing theory that SRPT is an
optimal algorithm for minimizing the mean response time
when the job size is known in advance \cite{proof}. For static requests, the size of the request (time required to 
service the request) can be well-approximated by the size of
the file requested by HTTP request, which is known to the
server. The SRPT based scheduling strategies uses this approximation for the scheduling of HTTP requests. The work
done to implement SRPT scheduling for web servers has
been done on both the application level and the kernel level. M. Crovella and R. Frangioso \cite{crovella} implemented SRPT 
connection based scheduling (priority to requests with smaller
size) at the application level and got improvement in the response time as compared to traditional PS scheduling but at
the cost of drop of throughput by some constant factor. Later, M. Harchol Balter \cite{balter}
implemented SRPT based scheduling strategy at the kernel
level and got better improvements in response time than in
\cite{crovella} and the throughput problem was eliminated. \\
\hspace*{5 mm}C. Murta \cite{fcf} introduced an extension to SRPT known
as FCF (Fastest Connection First) scheduling which takes
into consideration the network conditions instead of relying
only on file size of request. The strategy gives priority to
the requests with shorter size issued through faster connections.The information sharing between a Web server and
the TCP connections running in the server was taken into
consideration. They got improvements in the response time
as compared to SRPT, although SRPT provided the shortest server delay. They concluded that scheduling strategies
for web servers should take into consideration the network
conditions (WAN) and use them to set up priorities for the
requests. Ahmad AlSa'deh \cite{srrt} suggested SRRT (Shortest Remaining Response Time) scheduling for web servers which is
also an extension to SRPT. In addition to the file size, the
scheduling algorithm also takes into consideration current
RTT (Round-Trip Time) and TCP congestion window size
for the servicing of HTTP requests. They got an average
improvement of about 7.5\% over SRPT algorithm. For the
evaluation of results, \cite{srrt} did not take into account the variability of network bandwidth from user to user but instead
used same network bandwidth for all the clients throughout
the experiments. Further SRRT scheduling strategy was applicable only for static requests.\\
\hspace*{5 mm}Most of the papers have considered the idea of SRPT based scheduling policies \cite{crovella}, \cite{2006srpt}, 
\cite{fcf}, \cite{alpha}, \cite{2007srpt}, \cite{srrt}. But these scheduling strategies do not prevent the starvation of large size
requests in the presence of continuous arrival of short size requests. Further, all SRPT based policies are applicable only for the
static HTTP requests where the file size is known in advance. But today in most of the practical scenarios, web servers use dynamic
content as well (cgi and other non static requests). So, SRPT based policies cannot be applied in these scenarios.\\
\hspace*{5 mm}Bender, Chakravati and Muthukrishnan \cite{bender} proved that SRPT
will cause large files to have arbitrarily high response time.
This paper raised an important point that while choosing
a scheduling policy it is important to consider not only the
scheduling policy's performance but also whether the policy
is fair, i.e. some of the jobs have arbitrarily high response
time. There have been some research papers which studies the
types of unfairness caused by SRPT. For example, M. Gong
and C. Williamson \cite{gong} investigates about two different types
of unfairness (endogenous and exogenous) caused by Shortest Remaining Processing Time (SRPT) strategy.\\
\hspace*{5 mm} In order to overcome the problems related to SRPT based policies, in this work we present simple but effective 
scheduling policies with dynamic priority adjustment which addresses the problem of starvation and keeping the mean 
response time low, both simultaneously. Further, the policy will be able to handle the dynamic HTTP requests where the file size is 
not known in advance.

\subsection{Speed Up Scheduling}
\hspace*{5 mm}We present two scheduling algorithms named as \textit{SSU (Static Speed Up)} and \textit{DSU (Dynamic Speed Up)} for 
static (file size is known in advance) and dynamic environments (file size is not known in 
advance) respectively with a slight modification in the priority function defined in eq.\ref{Pfunction} based on implicit speed up. The algorithms are simple that assigns priority to the requests based on their specific characteristics.
The request with lowest value of priority function is accelerated to the server and thus, is chosen for next service. 
\subsubsection{Static Speed Up (SSU) scheduling:}
 This algorithm is non preemptive. Let $AT(r)$, $FS(r)$ and $LB(r)$ respectively denote the arrival time of 
request $r$, file size requested by $r$ and link bandwidth of the connection of the user sending request $r$ to the server i.e.
throughput of the user connection. The priority for request $r$ is assigned as follows:
\begin{equation}
\label{SSUequation}
Priority(r) = \frac {AT(r) \times FS(r)} {LB(r)} 
\end{equation}
\hspace*{5 mm}The request with minimum value of priority function is chosen next for service. If there are multiple such requests,
then the one with shortest file size is chosen. The idea is to give priority to small size requests issued through faster connection
in order to improve the mean response time but at the same time taking into consideration the arrival time of request in order to 
address the problem of starvation. The SSU algorithm is applicable only for static requests arriving at web server since
it includes the file size of request in its priority function.
\subsubsection{Dynamic Speed Up (DSU) scheduling:}
In order to schedule dynamic requests arriving at web server, we make
use of LAS (Least Attained Service) policy since the request size is not known in prior. LAS is a preemptive scheduling policy
which tries to predict the remaining job size by the amount of service, job has received so far and 
thus, mimics SRPT by giving preference to least attained service job. LAS will
create starvation for large size requests since arriving jobs will always have low value of attained service and thus, jobs existing in 
queue for long period of time will not get time to be serviced. Thus, we also take into consideration the arrival time of jobs to 
address this problem of starvation.\\
\hspace*{5 mm}Let $AT(r)$, $AS(r)$ and $LB(r)$ respectively denote the arrival time of request $r$, 
the attained service for $r$ (amount of time it has received service so far) and link bandwidth of the user connection 
sending request $r$ to the server.
Then the priority for request $r$ is assigned as follows:
\begin{equation}
\label{DSUequation}
 Priority(r) = \frac {AT(r) \times AS(r)}{LB(r)}
\end{equation}
The DSU scheduling algorithm is preemptive. We discuss the impact of the three parameters that we keep in the priority function on web requests :\\
\begin{itemize}
\item \textbf{Impact of Arrival Time in Priority:} The term $AT(r)$ in eq. \ref{SSUequation} and \ref{DSUequation} ensures that every request will
eventually be served. This is because AT (r) keeps on increasing for the incoming requests with respect to time and
we choose the request with minimum value of priority to get
next service. Thus, the time will always come when the request waiting for too long will have the minimum value of
priority function. Hence, the starvation for any request is
impossible.\\
\hspace*{5 mm} Consider a request $r_1$ with a large $FS(r_1)$, arriving at time $AT(r_1)$. Since requests keep coming, and arrival
time of requests keep increasing, there exists a request $r_2$ ,
such that $ \frac {AT(r_{2}) \times FS(r_2)}{LB(r_2)} > \frac{AT(r_1) \times FS(r_1)}{LB(r_1)}$, implying
that $r_1$ will be served before $r_2$ . That is, if $AT(r_2) > AT(r_1)$
and $\frac {AT(r_{1}) \times FS(r_1)} {LB(r_1)} < \frac {AT(r_2) \times FS(r_2)} {LB(r_2)}$ , $r_1$ is served before
$r_2$. Since remaining file sizes are finite and decreasing, and link bandwidth is also constant, each request will definitely
be served as priority of new requests keep increasing as their
arrival time is increasing.\\
\item \textbf{Impact of file size in Priority:} If there are various requests having near about similar
arrival time issued through the same link bandwidth connection, then the request with least file size is preferred over
others (like SRPT) in order to improve the response time.\\
\hspace*{5 mm}Consider a small size request $r_1$ with very small value of
$FS(r1)$ arrived at time $AT(r_1)$. Let there be another request $r_2$ already present in the request queue such that $AT(r_2)
< AT(r_1)$ and $FS(r_2)$ is much larger than $FS(r_1)$. If $\frac {AT(r_1) \times FS(r_1)}{LB(r_1)} < \frac{AT(r_2) \times FS(r_2)}{LB(r_1)}$, then $r_1$ will be served
before $r_2$ . It indicates that smaller value of $FS(r_1)$ (resulting in lower priority function value for $r_1$ than $r_2$ ) results
in preferring $r_1$ over $r_2$ for next service. Each request $r$ will
get served because there will be certain arrival time $AT(r\prime)$
(for some request $r\prime$ ) such that all the requests after that
irrespective of their file size ($>$ 1) will have higher value of
priority function than $r$.\\
\item \textbf{Impact of Link Bandwidth in Priority:} Link Bandwidth plays a very crucial role in influencing
the response time for web clients. Greater the link bandwidth, lesser is the response time. If there are set of requests 
present in the request queue having similar arrival time
and near about same file size, then the one with the fastest
link bandwidth is preferred over others which will result in
lesser response time.\\
\hspace*{5 mm}Consider a request $r_1$ with arrival time $AT(r_1)$ issued
through fast connection of say 100 Mbps. Then another request $r_2$ arrives with smaller value of $FS(r_2)$ as compared
to $FS(r_1)$ issued through a relatively slower connection of say 1 Mbps 
such that $\frac {AT(r_1) \times FS(r_1)}{LB(r_1)} < \frac {AT(r_2) \times FS(r_2)}{LB(r_2)}$ (because 
$LB(r_1)$ is much greater than $LB(r_2)$ ), then $r_1$ (fast
connection request) will be served before $r_2$ (slow connection
request) showing that how link bandwidth can impact the
value of priority function. Each request $r$ (even low bandwidth request) will get served irrespective of other requests
because there exist an arrival time beyond which their value
of priority function is higher than request $r$ assuming some
maximum link bandwidth (which is finite).
\end{itemize}
\subsection{Experiments and Results}
\hspace*{5 mm}We used a simple simulation model using trace-driven workload for the evaluation of results. We 
used a real one day logs of web requests arriving at International Institute Of Information Technology, Hyderabad
proxy server collected
on 9th September, 2012. Our workload consisted of a part of single day trace consisting of about 1 million requests. 
Each entry in the trace described a request made to the server and contained information about the request like arrival 
time of the request, URL requested, size of the request and other status information. 
The file size in our workload ranged from 56 bytes to 15 MB. \\
\hspace*{5 mm}We used a Scheduler Simulator in order to simulate all the requests arriving at a web server present in the 
trace based workload. Each request present in the trace can be considered as a process whose burst time 
is directly proportional to the file size requested and is inversely proportional to the link bandwidth of the request connection. 
The RAM of the system is 4GB and processor speed is 3.3GHz. The requests were simulated as per their information in the trace 
based workload using the Scheduler Simulator.\\
\hspace*{5 mm}For the purpose of assigning network bandwidth to each of the request, we partitioned the requests into following three types of classes :\\
\begin{itemize}
 \item \textit{Small} files (0 - 50 KB): These smaller sized requests contributed towards 64\% of the total workload.
 \item \textit{Medium}-sized files(50 KB - 500 KB): These requests contributed towards 32\% of the total requests present in our workload.
 \item \textit{Large} files ($>$ 500 KB): These requests contributed towards 4\% of the overall workload.
\end{itemize}
\hspace*{5 mm}We performed our experiments for two different types of extreme scenarios. \\
\begin{itemize}
 \item In \textbf{Scenario 1}, we assigned 1 Mbps, 10 Mbps and 100 Mbps of network bandwidth to \textit{Small},
\textit{Medium-sized} and \textit{Large} files respectively.
\end{itemize}
\begin{itemize}
 \item In \textbf{Scenario 2}, we assigned 1 Mbps, 10 Mbps and 100 Mbps of network bandwidth to \textit{Large},
\textit{Medium-sized} and \textit{Small} files respectively.
\end{itemize}
\subsubsection{Scenario 1:Results}
\begin{itemize}
 \item \textit{Small} files (0-50 KB) assigned to 1 Mbps connection.
 \item \textit{Medium-sized} (50 KB-500 KB)files assigned to 10 Mbps connection.
 \item \textit{Large} ($>$ 500 KB) files assigned to 100 Mbps connection.
\end{itemize}

\begin{table}
\centering 
\tbl{Overall mean and maximum response time (in sec) for both the scenarios for different scheduling strategies\label{websch-overall}.}{
\begin{tabular}{|p{1.5cm}|p{1.5cm}|p{1.5cm}|p{1.5cm}|p{1.5cm}|}
\hline
\multirow{2}{1.7cm}{Algorithm}  & \multicolumn{2}{|p{2cm}|}{\centering \textbf{\footnotesize{Scenario1}}} & \multicolumn{2}{|p{2cm}|}{\centering \textbf{\footnotesize{Scenario2}}} \\ \cline{2-5}
				&Mean&Max&Mean&Max\\
\hline
  PS&2.37&4.44&1.03&6.16\\ \hline
  SRPT&\textbf{1.58}&4.39&\textbf{0.5}&6.18\\ \hline
  SSU&\textbf{1.33}&\textbf{4.18}&\textbf{0.6}&\textbf{5.98}\\ \hline
  DSU&2.12&\textbf{4.26}&0.98&\textbf{6.08}\\ \hline
\end{tabular}}
\end{table}

\hspace*{5 mm}Table \ref{websch-overall} shows that both the SSU and DSU scheduling algorithms outperforms the existing scheduling algorithms
in terms of maximum response time validating our idea that these algorithms provide less scope for starvation. 
The reason behind this decrease, is that our scheduling algorithms take into consideration 
the arrival time (aging) of request which do not allow any request to stay for a very long period of time. SSU scheduling 
provides the minimum response time in Scenario 1 even better than SRPT scheduling. The reason behind 
this improvement is that SRPT prefers the shorter size requests independent of their slow network bandwidth (1 Mbps) 
connection in this scenario. However, the actual response time suffered by the client is influenced by both the size of the 
requested file and the link bandwidth. Greater the link bandwidth, lower is the response time.
SSU scheduling takes into consideration both the size of requested file as well as the link bandwidth of the user connection.
Note that, such improvements are not observed in DSU algorithm, because it does not know in prior the file size of request. It 
uses attained service measure for the estimation of file size. However, it outperforms PS scheduling (which also do not require
file size) in mean as well as max response time. The major advantage of DSU scheduling is that it can be used for the scheduling
of dynamic web requests (file size of request is not known in prior) whereas SRPT and SSU are not applicable in such cases.\\
\begin{table}
\centering 
\tbl{\textbf{Scenario 1:}Overall mean and maximum response time (in sec) for different types of requests for 
different scheduling strategies\label{websch1-detailed}.}{
\begin{tabular}{|p{1.70cm}|p{1.60cm}|p{1.60cm}|p{1.60cm}|p{1.60cm}|p{1.60cm}|p{1.60cm}|}
\hline
\multirow{2}{1.7cm}{Algorithm}&
\multicolumn{2}{|p{1.7cm}|}{\centering \textbf{\footnotesize{Small}}} & 
\multicolumn{2}{|p{1.7cm}|}{\centering \textbf{\footnotesize{Medium}}} &
\multicolumn{2}{|p{1.7cm}|}{\centering \textbf{\footnotesize{Large}}}\\ \cline{2-7}
				&Mean&Max&Mean&Max&Mean&Max\\
\hline
  PS&2.26&4.41&2.62&4.44&1.38&4.39\\ \hline
  SRPT&\textbf{0.64}&2.71&3.13&4.22&4.3&\textbf{4.39}\\ \hline
  SSU&1.23&4.10&1.55&4.18&0.77&\textbf{3.03}\\ \hline
  DSU&2.04&4.19&2.59&4.26&1.67&\textbf{4.13}\\ \hline
\end{tabular}}
\end{table}
\hspace*{5 mm}Table \ref{websch1-detailed} shows the mean and maximum response time for the different types of requests for different
scheduling strategies for Scenario 1. The small size requests get a very low mean response time for SRPT (because of it's tendency to 
favor small requests), whereas on the other hand, large size requests (connected through 100 Mbps link bandwidth) 
suffer from very high mean response time of 4.3 seconds by SRPT. SRPT only takes into consideration the remaining size of requested file 
thereby making the large size jobs to wait for too long. Such high penalization for large requests is not observed for 
both SSU and DSU scheduling because along
with the file size, they also consider the arrival time in priority to avoid starvation.\\ 
\hspace*{5 mm}From the above results, it can be concluded that in Scenario 1, SSU strategy 
outperforms SRPT  as well as PS scheduling in terms of mean response time without creating starvation like SRPT. DSU
scheduling also provides less scope of starvation and outperforms PS scheduling with a major advantage of its 
applicability in serving of dynamic web requests.
\subsubsection{Scenario 2:Results}
\begin{itemize}
 \item \textit{Small} files (0-50KB) assigned to 100Mbps connection.
 \item \textit{Medium-sized} (50-500KB) assigned to 10Mbps connection.
 \item \textit{Large} files ($>$ 500KB) assigned to 1Mbps connection.
\end{itemize}

\begin{table}
\centering 
\tbl{\textbf{Scenario 2:}Overall mean and maximum response time (in sec) for different types of requests for 
different scheduling strategies\label{websch2-detailed}.}{
\begin{tabular}{|p{1.70cm}|p{1.60cm}|p{1.60cm}|p{1.60cm}|p{1.60cm}|p{1.60cm}|p{1.60cm}|}
\hline
\multirow{2}{1.7cm}{Algorithm}&
\multicolumn{2}{|p{1.7cm}|}{\centering \textbf{\footnotesize{Small}}} & 
\multicolumn{2}{|p{1.7cm}|}{\centering \textbf{\footnotesize{Medium}}} &
\multicolumn{2}{|p{1.7cm}|}{\centering \textbf{\footnotesize{Large}}}\\ \cline{2-7}
				&Mean&Max&Mean&Max&Mean&Max\\
\hline
  PS&0.12&0.61&2.18&3.42&5.65&6.16\\ \hline
  SRPT&0.006&0.09&1.05&2.9&4.48&\textbf{6.18}\\ \hline
  SSU&0.07&0.2&1.17&2.9&4.4&\textbf{5.98}\\ \hline
  DSU&0.27&0.68&2.12&3.88&5.4&\textbf{6.08}\\ \hline
\end{tabular}}
\end{table}

\hspace*{5 mm} The starvation caused by SRPT for large files 
(connected through 1 Mbps bandwidth) is visible from the Table \ref{websch-overall} and \ref{websch2-detailed} for 
Scenario 2.
SSU and DSU scheduling performs better than SRPT scheduling in terms of maximum response time indicating the reduction
 in starvation. However, in this scenario the overall mean response time is minimum for SRPT better than SSU as well as
 DSU scheduling. This is because in this scenario, smaller size requests are connected 
 through faster link bandwidth connection. Thus, by preferring only the smaller size requests, 
 SRPT is also indirectly giving preference to faster connection requests (because smaller size requests are 
 connected through faster connections). SSU scheduling gives preference to both shorter size requests 
 as well as earlier arrived requests. Thus, it's response time is found to be
slightly lower than SRPT. In this case, SRPT provides an optimal response time. We can say that 
this scenario is the best case scenario for SRPT scheduling.
\subsection{Discussion}
\hspace*{5 mm}We have considered two extreme case scenarios for the 
variability in network bandwidth for different users. In first scenario, \textit{smaller}
size requests and \textit{larger} size requests are respectively connected through slower and faster link bandwidth 
connection while in the second scenario vice-versa.\\
\hspace*{5 mm}In \textbf{Scenario 1}, SSU scheduling outperforms both PS and SRPT scheduling.
SRPT scheduling blindly gives preference to shorter size requests irrespective of taking into account 
their slower network bandwidth in this scenario. It leads to the starvation of large size requests 
which are connected through faster link bandwidth connection. SSU scheduling takes into consideration 
both the size as well as link bandwidth thereby, improving the mean response time from 1.53 to 1.3 sec.
SSU scheduling reduces the problem of starvation in terms of maximum response time 
as compared to SRPT scheduling from 4.39 to 4.18 sec because it takes into consideration the arrival time of 
request and thus, prevents the request from waiting for too long. DSU scheduling outperforms PS scheduling in terms of mean
response time as well as max response time. DSU scheduling can further be applied for dynamic environments whereas SSU
and SRPT cannot.\\ 
\hspace*{5 mm} In \textbf{Scenario 2}, \textit{small} size requests are connected 
through faster connection while the \textit{large} size requests are connected
through slower connection. SRPT scheduling while giving preference to \textit{small} size requests 
is also indirectly giving preference to the requests with faster
connection (because of the nature of this scenario). 
Thus, SRPT provides an optimal response time in this scenario (best-case scenario for SRPT). 
SSU scheduling suffers a drop from 0.5 sec (as in SRPT) to 0.6 sec in terms of mean response time
as compared to SRPT scheduling. However, SSU scheduling again reduces the problem of 
starvation from 6.18 to 5.98 sec in terms of maximum response time as compared to SRPT. DSU scheduling
outperforms PS scheduling in terms of mean as well as maximum response time in this scenario as well. The reason behind
this improvement is that PS scheduling do not give any preference to short size requests however, DSU tries to mimic SRPT
by giving preference to least attained serviced request.
\begin{table}
\centering
\tbl{Scenario 1: Percentage of speeded up and slowed down requests for different scheduling strategies\label{websch1-SU}.}{
\begin{tabular}{|c|c|l|} \hline
Strategy&\% of achieved speed-up &\% of slow down\\ \hline
PS&49.8\%&50.2\% \\ \hline
SRPT&61.2\%&37.6\% \\ \hline
SSU&62.6\%&37.2\% \\ \hline
DSU&51.2\%&48\% \\ \hline
\end{tabular}}
\end{table}

\begin{table}
\centering
\tbl{Scenario 2: Percentage of speeded up and slowed down requests for different scheduling strategies\label{websch2-SU}.}{
\begin{tabular}{|c|c|l|} \hline
Strategy&\% of achieved speed-up &\% of slow down\\ \hline
PS&85.25\%&14.75\% \\ \hline
SRPT&91.75\%&8\% \\ \hline
SSU&91\%&7.5\% \\ \hline
DSU&86.25\%&13.75\% \\ \hline
\end{tabular}}
\end{table}
\hspace*{5 mm} Table \ref{websch1-SU} and \ref{websch2-SU} indicates the percentage of speeded up and slowed down requests \textit{w.r.t.} FCFS scheduler
for Scenario $1$ and $2$
respectively. SSU and SRPT scheduling dominates in terms of \% of achieved speed up as compared to other scheduling algorithms for both the scenarios.
DSU performs better than PS scheduling in both the scenarios.\\
\subsection{User Abort Analysis}
\hspace*{5 mm}For starvation analysis, we used \textit{Maximum Response Time} metric as a measure of unfairness till now. Now, we are going to analyze starvation
for the different scheduling strategies using another metric, \textit{i.e.} User Abort.\\
\hspace*{5 mm}Today, users cannot wait for large amount of time for getting their request serviced. They might look for other alternatives 
if some particular website makes them wait for too long. Thus, we may assume that the user waits for certain specific amount of 
time (say \textit{user threshold}) till his request gets served after which
he aborts. This gives rise to a situation known as User Abort.\\
\hspace*{5 mm}We set the value of user threshold to be slightly less than the maximum response time for unbiased PS 
scheduling and measure the percentage of users having response time greater than user threshold (percentage of user abort) for the PS, SRPT, SSU and DSU scheduling strategies.\\
\begin{table}
\centering
\tbl{Percentage Of User Abort in two scenarios for different scheduling strategies\label{userabort}.}{
\setlength{\tabcolsep}{18pt}
\begin{tabular}{|c|c|l|} \hline
Strategy&Scenario 1&Scenario 2\\ \hline
PS&3.1\%&2.1\% \\ \hline
SRPT&3\%&1.8\%\\ \hline
SSU&\textbf{0.9\%}&\textbf{0.25\%} \\ \hline
DSU&\textbf{1.6\%}&\textbf{1.1\%} \\ \hline
\end{tabular}}
\end{table}
\hspace*{5 mm} Table \ref{userabort} clearly indicates that SRPT penalizes the large size requests whereas on the other hand, 
SSU and DSU, both strategies provide less scope of starvation thereby handling more number of users. SSU and DSU
scheduling outperforms both PS and SRPT in terms of user aborts and thus, \textit{these algorithms ensures more reliable 
service than either PS or SRPT}.
\section{Applications: CPU Scheduling}
\hspace*{5 mm}We extend our priority functions designed for implicit speed up for the purpose of CPU Scheduling and demonstrate the 
effectiveness of our policies using simulation based model.
\subsection{Introduction}
\hspace*{5 mm}In real time uniprocessor multiprogrammed systems, only one process can run at a time. All the processes which are waiting
for CPU resources and are ready to execute resides in the ready queue. Whenever CPU becomes idle (either because executing process
is waiting for I/O routine to complete or process terminates), a new process needs to be selected for execution among all the processes present
in the ready queue. The task of selecting a new process among all the runnable processes present in the ready queue for
execution is referred to as \textit{CPU Scheduling}. The module that is responsible for giving control of the CPU resources
to the selected process by \textit{CPU Scheduler} is referred to as \textit{dispatcher}. CPU Scheduling is a 
fundamental problem in terms of minimizing the mean wait time, ensuring fairness among processes and enhancing the overall
performance of the system.
\subsection{Overview of existing CPU Scheduling Algorithms}
\hspace*{5 mm}Many CPU scheduling algorithms such as FCFS, Shortest Job First, Round-Robin, Priority Scheduling etc. have already been proposed.
First Come First Serve (FCFS) scheduling selects a job with least arrival time. This algorithm is fair to all and will not create starvation for any of the job.
However in FCFS, the earlier arrived process with large duration time may unnecessarily delay the short jobs arrived later
leading to high mean wait time. This phenomena is referred to as Convoy Effect. Shortest Job First algorithm selects a process
with least duration time among all the runnable processes. SJF algorithm is provably optimal in giving minimal average wait time
for all the processes \cite{proof}. But the tendency of SJF algorithm to prefer shorter jobs might create starvation for large jobs.
Thus, SJF algorithm might not be fair towards long processes in the presence of continuous arrival of short processes.
Further, practically process duration time is not known in advance and thus, SJF algorithm cannot be applied 
directly for the purpose of CPU scheduling. There are heuristics like \textit{exponential average} 
which predicts the next CPU burst time of a process depending upon its previous CPU burst time. Round Robin scheduling is similar
to FCFS scheduling but preemption is added to allow switching between processes. In RR algorithm, each of the process 
is given a fixed time slice known as \textit{time quantum}. The process runs until the time quantum expires, then context
switch takes place leading to the execution of the next arrived process and so on. Very large value of time 
quantum will make Round Robin exactly similar to FCFS algorithm. If the value of time quantum is too less, then
most of the time would be spent in context switching between the processes. The average wait time under RR policy is often long.
Priority based scheduling selects a process with highest priority to schedule next. This scheduling technique may create
starvation for low priority processes if high priority processes keep coming with respect to time.\\
\hspace*{5 mm}Wait time and Starvation (situation of waiting for indefinite time) are two of the most important factors for choosing a CPU Scheduling policy for
processes. We present two Speed Up scheduling techniques for the purpose
of CPU Scheduling which aims at reducing mean wait time but at the same time ensuring less scope of starvation - \textit{Static Speed Up Process Scheduling (SSUPS)} (assumes process burst time is known) and \textit{Dynamic
Speed Up Process Scheduling (DSUPS)}
(process burst time is not known in advance).
\subsection{Speed Up Process Scheduling Algorithms}
\subsubsection{Static Speed Up Process Scheduling (SSUPS)}This scheduling technique is non-preemptive. The priority for a process
$p$ is assigned as follows:
\begin{equation}
 Priority(p) = AT(p) * BT(p)
\end{equation}
where $AT(p)$ denotes the arrival time of process $p$ and $BT(p)$ denotes the CPU burst time of process $p$. The process
with least value of priority is chosen to schedule next. If there are multiple processes with minimum priority value, then 
the process with smallest burst time is chosen. This
scheduling technique assumes that process CPU burst time is known in advance. The idea is to give preference to small
processes but this preference takes into consideration the arrival time to avoid starvation.
\subsubsection{Dynamic Speed Up Process Scheduling (DSUPS)}DSUPS scheduling technique is preemptive and is more practical than
SSUPS because it does not make use of CPU burst time to assign priority but instead use attained service parameter to
estimate the CPU burst time of the process. The priority for process $p$ is assigned as follows:
\begin{equation}
 Priority(p) = AT(p)*AS(p)
\end{equation}
where $AT(p)$ denotes the arrival time of process $p$ and $AS(p)$ denotes the attained service duration for process $p$.
The process with minimum priority value is chosen.
\subsection{Experiments and Results}
\hspace*{5 mm}We used a simulation based model for conducting experiments and getting results. We assumed that processes are arriving
in Poisson distribution and their burst times are exponentially distributed. The number of processes for experiments
were 10 thousand. We conducted experiments for different values of process load $\rho$ and present here the results for the
same. We also used real process logs dataset (PLD) used in Chapter 5 for comparison between different scheduling algorithms.
\begin{table}
\centering
\tbl{Mean Wait Time (in time units) for different CPU Scheduling policies based on different system load values.\label{cpu-mean-wait-time}}{
\setlength{\tabcolsep}{18pt}
\begin{tabular}{|c|c|c|c|c|c|l|} \hline
System load ($\rho$)&FCFS&SJF&SRPT&RR&SSUPS&DSUPS\\ \hline
 0.8&74&\textbf{37}&\textbf{27}&78&\textbf{38}&77\\ \hline
0.9&129&\textbf{57}&\textbf{45}&129&\textbf{59}&126\\ \hline
1&135&\textbf{60}&\textbf{49}&143&\textbf{63}&132 \\ \hline
1.1&672&\textbf{188}&\textbf{170}&628&\textbf{194}&600 \\ \hline
1.2&932&\textbf{259}&\textbf{248}&868&\textbf{273}&890 \\ \hline
1.3&1543&\textbf{443}&\textbf{436}&1451&\textbf{520}&1497 \\ \hline
PLD&3425&\textbf{987}&\textbf{951}&2819&\textbf{1122}&2911 \\ \hline
\end{tabular}}
\end{table}

\begin{table}
\centering
\tbl{Maximum Wait Time (in time units) for different CPU Scheduling policies based on different system load values.\label{cpu-max-wait-time}}{
\setlength{\tabcolsep}{18pt}
\begin{tabular}{|c|c|c|c|c|c|l|} \hline
System load ($\rho$)&FCFS&SJF&SRPT&RR&SSUPS&DSUPS\\ \hline
 0.8&\textbf{455}&\textbf{1448}&\textbf{1461}&1059&\textbf{1420}&1802\\ \hline
0.9&\textbf{589}&\textbf{1554}&\textbf{1713}&1417&\textbf{1520}&2397\\ \hline
1&\textbf{695}&\textbf{1823}&\textbf{1823}&1548&\textbf{1802}&2051 \\ \hline
1.1&\textbf{1796}&\textbf{7887}&\textbf{7894}&6759&\textbf{6453}&8380 \\ \hline
1.2&\textbf{2351}&\textbf{8979}&\textbf{8979}&7170&\textbf{6658}&8395 \\ \hline
1.3&\textbf{3371}&\textbf{10671}&\textbf{10671}&9233&\textbf{7262}&8774 \\ \hline
PLD&7053&\textbf{15431}&\textbf{16529}&15495&\textbf{13051}&\textbf{14186} \\ \hline
\end{tabular}}
\end{table}

\begin{table}
\centering
\tbl{Standard Deviation of wait time for different CPU Scheduling policies based on different system load values.\label{cpu-std-dev}}{
\setlength{\tabcolsep}{18pt}
\begin{tabular}{|c|c|c|c|c|c|l|} \hline
System load ($\rho$)&FCFS&SJF&SRPT&RR&SSUPS&DSUPS\\ \hline
0.8&\textbf{94}&\textbf{107}&\textbf{110}&144&\textbf{105}&224\\ \hline
0.9&190&\textbf{152}&\textbf{171}&250&\textbf{149}&351\\ \hline
1&197&\textbf{168}&\textbf{180}&268&\textbf{165}&349 \\ \hline
1.1&879&\textbf{781}&\textbf{832}&1125&\textbf{714}&1474 \\ \hline
1.2&1208&\textbf{1010}&\textbf{1018}&1474&\textbf{920}&1879 \\ \hline
1.3&1835&\textbf{1526}&\textbf{1536}&2080&\textbf{1311}&2154 \\ \hline
PLD&\textbf{2265}&2826&\textbf{2782}&3293&\textbf{2253}&3610 \\ \hline
\end{tabular}}
\end{table}
Experimental results show that SSUPS scheduling algorithm provides far better mean wait time as compared to FCFS and Round
Robin scheduling but slightly higher than SJF and SRPT (which are optimal algorithms for providing minimal mean wait time).
Further, our SSUPS algorithm reduces the problem of starvation in terms of maximum wait time as observed in SJF and SRPT
(preemptive version of SJF) for all values of system load since our algorithm takes into consideration the arrival time
to avoid large delay for any of the process. On the other hand, SJF and SRPT makes large size process to wait for too long
resulting in high value of maximum wait time.
Low value of maximum wait time guarantees that all the processes get good service. DSUPS algorithm is also successful in providing better mean wait time as compared to FCFS and Round Robin for majority of 
system load values and provides lesser maximum wait time as compared to SJF and SRPT under overload situation ($\rho>=1.2$).
FCFS algorithm provides the least maximum wait time since it is fair to all the processes but at the expense of high mean wait time.
By examining the experimental results from Table \ref{cpu-mean-wait-time} and \ref{cpu-max-wait-time}, we can conclude that our process scheduling algorithms (SSUPS and DSUPS) are successful in providing less mean wait time
but at the same time ensuring less scope of starvation. Thus, our proposed algorithms addresses the trade-off between wait time and starvation. SSUPS algorithm performs much better than DSUPS algorithm in terms of both mean as well as maximum wait time. 
The reason is that SSUPS algorithm assumes that process burst time is known in advance whereas DSUPS algorithm tries to 
estimate the process burst time using attained service parameter. But DSUPS algorithm is more practical than SSUPS algorithm
since process burst time is not known in advance.\\
\hspace*{5 mm}From \cite{osbook}, regarding interactive systems (such as time sharing systems), it is stated that minimizing
the \textit{variance} measure for wait time is more important than to minimize the \textit{mean} measure. A system with 
reasonable and predictable waiting time may be considered more desirable than a system that is faster on average but is highly variable.
Table \ref{cpu-std-dev} shows the value of standard deviation (square of standard deviation is variance) in wait time for different CPU scheduling algorithms based on different values of 
system load. We experimentally find that SSUPS algorithm provides the minimal variance in wait time and thus, \textit{outperforms all the existing algorithms in terms of variance minimization}. 
\subsection{Speed-up/Slow-down Analysis}
\hspace*{5 mm}Speed Up concept provides a new evaluation criteria for comparing different 
scheduling algorithms based on their speed-up/slow-down characteristics. We analyze the existing scheduling algorithms
from the perspective of speed up in this sub-section.\\
\hspace*{5 mm}Consider a set of processes $<p_1$, $p_2$, $p_3$ , ..... , $p_n>$. Let their corresponding wait times under
scheduling policy $X$ and $Y$ be denoted by $<w_{x_1}$, $w_{x_2}$, $w_{x_3}$ , ...... , $w_{x_n}>$ and
$<w_{y_1}$, $w_{y_2}$, $w_{y_3}$ , ...... , $w_{y_n}>$ respectively.
\begin{definition}
 A scheduling algorithm $X$ is said to speed up process $p_i$ w.r.t scheduling algorithm $Y$ iff $w_{x_i}-w_{y_i} < 0$.
 Similarly, $X$ is said to slow down process $p_i$ w.r.t. $Y$ iff $w_{x_i}-w_{y_i} > 0$.
\end{definition}
\begin{table}
\centering 
\tbl{Percentage of achieved speed up (SU) and slow down (SD) for different CPU scheduling strategies with respect to other 
scheduling strategies\label{cpusch-SU1}.}{
\begin{tabular}{|p{1.70cm}|p{1.6cm}|p{1.65cm}|p{1.65cm}|p{1.65cm}|p{1.65cm}|p{1.65cm}|}
\hline
\multirow{2}{1.7cm}{Algorithm}&
\multicolumn{2}{|p{1.7cm}|}{\centering \textbf{\footnotesize{FCFS}}} & 
\multicolumn{2}{|p{1.7cm}|}{\centering \textbf{\footnotesize{RR}}} &
\multicolumn{2}{|p{1.7cm}|}{\centering \textbf{\footnotesize{SJF}}} \\ \cline{2-7}
				&SU&SD&SU&SD&SU&SD\\
\hline
  FCFS&0\%&0\%&34.8\%&61.4\%&9.4\%&\textbf{82.8}\% \\ \hline
  RR&61.4\%&34.8\%&0\%&0\%&8.8\%&\textbf{87}\% \\ \hline
  SJF&\textbf{82.8\%}&9.4\%&\textbf{87\%}&8.8\%&0\%&0\% \\ \hline
  SRPT&\textbf{88\%}&8\%&\textbf{93.8\%}&1.8\%&\textbf{75.4}\%&14.4\% \\ \hline
  SSU&\textbf{81.6\%}&10.6\%&\textbf{87.2\%}&8.2\%&16.6\%&42.2\% \\ \hline
  DSU&\textbf{72.8\%}&24.4\%&\textbf{76.8\%}&19.6\%&\textbf{48\%}&\textbf{47.8\%} \\ \hline
\end{tabular}}
\end{table}
\begin{table}
\centering 
\tbl{Percentage of achieved speed up (SU) and slow down (SD) for different CPU scheduling strategies with respect to other 
scheduling strategies\label{cpusch-SU2}.}{
\begin{tabular}{|p{1.70cm}|p{1.65cm}|p{1.65cm}|p{1.65cm}|p{1.65cm}|p{1.65cm}|p{1.65cm}|}
\hline
\multirow{2}{1.7cm}{Algo.}&
\multicolumn{2}{|p{1.7cm}|}{\centering \textbf{\footnotesize{SRPT}}} & 
\multicolumn{2}{|p{1.7cm}|}{\centering \textbf{\footnotesize{SSU}}} &
\multicolumn{2}{|p{1.7cm}|}{\centering \textbf{\footnotesize{DSU}}}\\ \cline{2-7}
				&SU&SD&SU&SD&SU&SD\\
\hline
  FCFS&8\%&88\%&10.6\%&81.6\%&24.4\%&72.8\% \\ \hline
  RR&1.8\%&93.8\%&8.2\%&87.2\%&19.6\%&76.2\% \\ \hline
  SJF&14.4\%&75.4\%&42.2\%&16.6\%&47.8\%&48\% \\ \hline
  SRPT&0\%&0\%&\textbf{78.2\%}&13.8\%&\textbf{60.8\%}&8.8\% \\ \hline
  SSU&13.8\%&78.2\%&0\%&0\%&47.4\%&48.6\% \\ \hline
  DSU&8.8\%&60.8\%&48.6\%&47.4\%&0\%&0\% \\ \hline
\end{tabular}}
\end{table}
\hspace*{5 mm}Table \ref{cpusch-SU1} and \ref{cpusch-SU2} presents the comparison analysis for different CPU Scheduling algorithms
based on their speed-up/slow-down characteristics for $\rho = 1.1$. Round Robin speeds up $61.4\%$ of the processes w.r.t. FCFS scheduling.
All the four scheduling algorithms SJF, SRPT, SSUPS and DSUPS speeds up nearly 70-90\% of the processes w.r.t. FCFS and Round
Robin Scheduling. SRPT scheduling speeds up majority of the processes w.r.t. any other scheduling strategy. SSUPS scheduling
w.r.t. DSUPS scheduling speeds up 47.8\% processes and slows down 48.6\% processes. In this manner, we can compare between
any two scheduling algorithms based on their speedup/slowdown characteristics.
\subsection{Discussion}
\hspace*{5 mm}Speed Up concept provides a new mechanism to compare
and contrast different scheduling algorithms. Speed Up can be implemented by various techniques. In \cite{kafeza} and \cite{wecwis}, it was implemented by
location table and swapping of jobs. In this paper, we use priority function to seamlessly and certainly perform speed up. Any scheduling
algorithm that uses ranking function to select next job can potentially speed up that job.\\
\hspace*{5 mm}From speed up perspective, we have assumed that no process is requesting for speed up. We extended our implicit priority
functions of speed up problem for the purpose of CPU Scheduling. Experimental results show that our proposed techniques
are successful in reducing the mean wait time and at the same time providing less scope of starvation, thus addressing
the trade-off between wait time and starvation. Further, SSUPS algorithm provides minimal variance in wait time as compared to all the other existing scheduling algorithms.
\section{Conclusion and Future Work}
\hspace*{5 mm}Today in flexible and dynamic application environments, user might request for faster execution of some already executing instances. 
In such cases, the system should be able to respond to such on-line requests for speed up. The problem of scheduling of speed up
requests at run-time has not been adequately studied in literature before. In this paper, we modeled the 
Speed Up Scheduling problem without acquiring additional resources to handle on-line speed up requests and 
analyzed two different aspects of it - Speed Up with no constraints and Fine Grained Selective Speed Up. 
We provided efficient implicit techniques to address speed up problem where the notion of 
speed up is incorporated in the priority function. Experimental results show that our proposed algorithms are able to provide
almost the similar achieved speed up as compared to existing speed up algorithms. Further, our speed up algorithms outperforms
the existing algorithms in terms of slow down caused to non-urgent jobs thereby providing remedies for the delayed jobs
alike existing algorithms where speed up is achieved for urgent jobs at the expense of high slow down for the rest of the jobs. Overall mean wait time is improved dramatically by
our speed up algorithms owing to the design of the priority function. We provided GPSU algorithm to address more specific
case of speed up problem - Fine Grained Selective Speed Up (each of the urgent job could request for specific percentage of
speed up at run-time) which has not been yet addressed. Our algorithms are computationally efficient than existing speed up
algorithms where there is an overhead of maintaining a location table.\\
\hspace*{5 mm}For web scheduling, the performance of SRPT degrades dramatically in an environment where there is a 
high variability in link bandwidth. Using our implicit techniques for speed up problem, we provided two web scheduling
algorithms SSU and DSU for static and dynamic environments respectively. We established the usefulness of our algorithms with
a simulation based model using trace driven workload. We extended our algorithms for CPU Scheduling and experimentally finds
that our algorithms are successful in reducing the mean as well as variance in wait time and at the same time providing less scope for
starvation. \\
\hspace*{5 mm} In this work, we provided a framework for addressing speed up related problems. But still there are 
several open issues that need to be addressed. Throughout this work, we assumed that no additional resources could be
acquired. An extension of our work could be to examine how such on-line requests for speed up 
can be handled if multiple servers and other additional resources are available so as to maximize the number of
speeded up urgent jobs but at the same time ensuring efficient resource utilization. We also assumed that jobs are
independent of dependence constraints throughout this work. It would be interesting to see how
we can speed up jobs requesting for it where that job is either dependent on some other non-urgent job
or on another job requesting for speed up ? In the presence of multiple queues, it would be interesting to see how 
load balancing could be used in achieving speed up.\\\\\\

\bibliographystyle{ACM-Reference-Format-Journals}
\bibliography{sigproc}



\end{document}